%% file: main.tex
\documentclass[fleqn,usenatbib]{mnras}

\usepackage[T1]{fontenc}

\DeclareRobustCommand{\VAN}[3]{#2}
\let\VANthebibliography\thebibliography
\def\thebibliography{\DeclareRobustCommand{\VAN}[3]{##3}\VANthebibliography}

\usepackage{graphicx}	
\usepackage{amsmath}	
\usepackage{amssymb}	
\usepackage{tabularx} 
\usepackage{multirow}
\usepackage{subcaption}
\usepackage{xcolor}\usepackage{color,soul}

\usepackage{newtxtext,newtxmath}

\title[The Eccentric Laplace Surface]{Laplace surface dynamics, revisited: satellites, exo-planets and debris with distant, eccentric companions}
\author[M. Farhat and J. Touma]{
Mohammad A. Farhat,$^{1}$\thanks{ mohammad.farhat@obspm.fr}
Jihad R. Touma$^{2, 3}$\thanks{jt00@aub.edu.lb}
\\
\\
$^{1}$IMCCE, CNRS, Observatoire de Paris, PSL University, Sorbonne Université, 77 Avenue Denfert-Rochereau, 75014, Paris, France\\
$^{2}$Department of Physics, American University of Beirut
PO Box 11-0236, Riad El-Solh, Beirut 11097 2020, Lebanon\\
$^{3}$Center for Advanced Mathematical Sciences, American University of Beirut, PO Box 11-0236, Riad El-Solh, Beirut 11097 2020, Lebanon
}
\date{Accepted XXX. Received YYY; in original form ZZZ}

\pubyear{2021}

\begin{document}
\label{firstpage}
\pagerange{\pageref{firstpage}--\pageref{lastpage}}
\maketitle

\begin{abstract}
To date, studies of \textit{Laplace Surface} dynamics have concerned themselves with test particle orbits of fixed shape and orientation in the combined field of an oblate central body (to which the particle is bound) and a distant, inclined, companion which is captured to quadrupolar order. While amply sufficient for satellites around planets on near-circular orbits, the quadrupolar approximation fails to capture essential dynamical features induced by a wide binary companion (be it a star, a planet or a black hole) on a fairly eccentric orbit. With similar such astronomical settings in mind, we extend the classical Laplace framework to higher multipoles, and map out the backbone of stationary orbits, now complexified by the broken axial symmetry. Eccentric and inclined Laplace equilibria, which had been presaged in systems of large enough mutual inclination, are here delineated over a broad range of mutually inclined perturbations.  We recover them for test particles in the field of a hot Jupiter and a wide eccentric stellar binary, highlighting their relevance for the architecture of multi-planet systems in binaries. We then extend and deploy our machinery closer to home, as we consider the secular dynamics of Trans-Neptunian Objects (TNOs) in the presence of a putative ninth planet. We show how generalized Laplace equilibria seed islands for Trans-Neptunian objects to be sheltered around, islands within chaotic seas which we capture via Poincar\'{e} sections, while highlighting a beautiful interplay between Laplace and Kozai-Lidov secular dynamical structures. An eminently classical tale revived for the exo-planetary 21st century!
\end{abstract}

\begin{keywords}
gravitation -- celestial mechanics -- planets and satellites: dynamical evolution and stability -- Kuiper belt: general

\end{keywords}


\input{Introduction}

\input{Secular_Dynamics}
\input{P9}

\input{Discussion}
\input{Conclusion}

\section*{Acknowledgements}
 We thank Mher Kazandijian and Antranik Sefilian for valuable assistance with the modal analysis toolbox. 

\section*{Data Availability}
The data underlying this article pertaining to the orbital parameters of Trans-Neptunian Objects are collected from the IAU Minor Planet Center at [https://www.minorplanetcenter.net/iau/mpc.html].
 



\bibliographystyle{mnras}
\bibliography{References} 


\appendix
\input{A1}

\bsp	
\label{lastpage}
\end{document}

%% file: Introduction.tex
\section{Introduction} 

The orbital architecture of planetary satellites provides the ideal astronomical setting in which to explore hierarchical dynamical processes, combining the effect of planetary oblateness on the inside, and the solar tide on the outside. In that setting, the existence of a “proper” inclination, at which the circular orbit of a satellite around an oblate and oblique planet is stationary, was recognized by \citet{laplace1805vol} in his classic study of Jupiter’s satellites. That peculiar inclined plane, which is now referred to as the Laplace plane, approaches the planet’s equator on the inside and tends to the planet’s orbit around the Sun with increasing orbital radius. This single parameter family of Laplace planes is enveloped by a surface which interpolates as it warps between the equator and the orbit of the planet. It is naturally referred to as the \textit{Laplace Surface}, an idealized object of persistent fascination, which has structured our understanding of the formation and evolution of planetary satellites, and accompanying ring systems for more than two hundred years now.

With a view to greater realism, and consequent complexity, Tremaine, Touma, and Namouni (\cite{Tremaine}, TTN hereafter) relaxed the assumption of circular equilibria, and studied the stability of the classic \textit{Laplace Surface} to perturbations in a satellite's eccentricity. They thus opened a can of worms, and left the field with a series of novel results, touching primarily on highly oblique planets, and including a hitherto unsuspected instability, and the likelihood of stints of chaotic evolution in the course of orbital migration. Their study foresaw, and ushered a stream of exo-planetary applications, and generalizations allowing for: warps of disks in binaries \citep{charnoz_canup_crida_dones_2018}, warps in circumplanetary disks \citep{zanazzi2016extended}, then non-gravitational perturbations such as radiation pressure on dust grains \citep{rosengren2014laplace, tamayo2013dynamical}. The circle was recently closed back onto the solar system, when it was realized that origin around a fast spinning and oblique Earth, which is favored on geochemical grounds,  could generate a Moon whose Laplace plane may very well be prone to TTN’s eccentricity instability! \citep{cuk2016tidal, tian2020vertical, cuk2021tidal}

The present work begins where TTN left the \textit{Laplace Surface}, and relaxes yet another assumption in the original Laplace story by explicitly accounting for an eccentric binary companion, thus effectively breaking the hitherto assumed axisymmetry of the outer perturber. This level of generality is again demanded by exo-planetary systems, the disks that generate them, remnant debris disks, all in the presence of a massive wide binary companion on an eccentric (and inclined) orbit. It is a remarkable result of secular celestial mechanics that the symmetry breaking effect of eccentricity appears first at the octupolar order, and it is at and above that order that our calculations will be conducted.

Our formalism and associated results are of broad applicability, over a range of systems and scales within them. Following general results on the classical Laplace surface and its fate, we illustrate securely hierarchical regimes with a multiplanet system in a wide eccentric binary. We then overcome the breakdown of hierarchy as we explore the peculiar orbital architecture of trans-Neptunian objects (TNOs) in the presence of a putative ninth planet \citep{trujillo2014sedna, batygin2019planet}. In so doing, we generalize the planar structure studied by \cite{beust2016orbital} then \cite{saillenfest2017non} to mutually inclined perturbers, as we map families of inclined, eccentric, Laplace equilibria in the presence of solar system giants (inner quadrupole forcing) and an eccentric inclined binary companion (hypothesized ninth planet). Families of equilibria and their intricate bifurcations provide the desired skeletal structure with which to properly explore this intriguing scenario, and variations around it, whether allowing for the combined effect of secular and mean motion resonances \citep{malhotra2016corralling}, or the combined effect of a distant inclined planetary core and a self-gravitating debris disk \citep{silsbee2018producing,sefilian2019shepherding}. 

The context lends itself naturally to an interpolation between relatively strong inner quadrupolar forcing (and associated outer Kozai-Lidov dynamics), then relatively strong outer multipolar forcing (and associated eccentric Kozai-Lidov cycling) with the unfolding of Laplace equilibria in between. But we are getting ahead of ourselves! So without further ado, we shall start where we must, with the Hamiltonian governing the orbit averaged dynamics of the infamous test particle in the elegant vectorial formalism.

%% file: Secular_Dynamics.tex
\section{Laplace Surface Dynamics: Model and Variations}

We are interested in the motion, specifically the relative orbital equilibria, of a test particle in the combined gravitational field of a central point particle, and an inner (quadrupolar) then outer (octupolar) perturber, as reflected in the Hamiltonian $H_p = H_K + \Phi_i + \Phi_{o}$ with

\begin{equation}
    H_K= \frac{1}{2}\Dot{\vec{r}}^2 -\frac{G M}{r}, 
\end{equation}

\begin{align}
\nonumber
   \Phi_i(r,\theta) &= \frac{GM Q_2}{r^3}P_2(\cos\theta)\\
   &=\frac{GM Q_2}{2r^5}\bigg[3(\vec r\cdot\hat{n}_p)^2 - r^2\bigg],
\end{align}

and

\begin{align}\nonumber
    \Phi_{o}(\vec{r},\vec r^{\prime})=\frac{-GM_{o}}{r^{\prime}}\bigg[&1+\frac{\vec{r}\cdot\vec{r}^{\prime}}{r^{\prime2}}-\frac {r^2}{2r^{\prime 2}}+\frac{3(\vec{r}\cdot\vec{r}^{\prime})^2}{2r^{\prime 4}}+\frac{5}{2}\frac{(\vec{r}\cdot\vec{r}^{\prime})^3}{r^{\prime 6}}\\
    &-\frac{3}{2}\frac{\vec{r}\cdot\vec{r}^{\prime}}{r^{\prime 4}} r^2+...\bigg].
\end{align}
Under $H_K$ alone, a bound test particle, with position vector $\vec{r}$ and velocity vector $\Dot{\vec{r}}$, evolves on a spatially fixed Keplerian ellipse in the central force field of a Newtonian point particle of mass $M$. We do not have reasons to consider general relativistic (read Post-Newtonian) corrections in this work, though it should be quite straightforward and interesting to explore their signature in future instalments.

Inner to the test particle's orbit, $\Phi_i (\vec{r}, \hat{n}_p)$ models a non-spherical mass distribution, here captured up to quadrupolar order, essentially an axisymmetric gravitational perturbation of strength $Q_2$ and axis of symmetry $\hat{n}_p$. Outside that same orbit, we envisage a point particle of mass $M_o$ bound to the central body with position vector $\vec{r}^{\prime}$ (on a fairly eccentric Keplerian orbit) and perturbing the test particle with potential $\Phi_{o}(\vec r, \vec r^\prime) = -GM_o /|\vec r -  \vec r ^\prime | $, here captured to octupolar order.

Classically, the central body stands for the monopolar contribution of a planet (Jupiter say), and the inner quadrupole for that planet's dominant departure from spherical symmetry, namely the effect of its equatorial bulge. The strength $Q_2 = J_2 R_p^2$ of that bulge ($R_p$ being the planet's radius and $J_2$ its gravitational second zonal harmonic) can be further amended with the contribution of a system of inner satellites with mass and semi-major axis $\{m_i, a_i\}, i=1\dots n$:
\begin{equation}
    Q_2'= Q_2 +\frac{1}{2}\sum^n_{i=1} \frac{a_i^2m_i}{M}, 
\end{equation}
here all assumed revolving on circular orbits, in the planet's equatorial plane. 

The outer body is then typically the star hosting the planetary system in question (say the Sun), and its perturbative effect in a hierarchical architecture is often limited to the quadrupolar order, at which any orbital eccentricity appears as a parameter affecting the strength of perturbation to that order (more on that below). 

This, in particular, is the context in which Laplace and followers worked, the same model around which TTN elaborated their generalization to eccentric Laplace equilibria. That context was further applied to a range of astrophysical settings, allowing for variations on the central body (a host star in a wide binary system), the inner quadrupole (a hot Jupiter, a system of planets), the outer perturber (a binary companion, a distant planet) and the test particle (a planet in a hierarchical arrangement, a particle in a debris disk). 

By accounting for higher order contributions from the outer perturber, we take a first, and already quite challenging step, towards breaking the implicit axisymmetry of the classical model. The aim is to identify the skeletal structure of surviving  eccentric, inclined, Laplace equilibria, when sufficient allowance is made for the outer perturber's lopsided mass distribution. 

To get anywhere close to recovering those novel equilibria, we will need to proceed, as others did before us, by  averaging the model Hamiltonian, with all the caveats associated with that averaging, whether allowing for orbital resonances, or contributions from higher order effects, in mildly hierarchical systems. It is our belief that the skeleton that we shall map out will provide a secure scaffolding for dynamical insights, then further generalizations, and modifications. 

With a view to developments that follow,  we introduce an orbital reference frame with the following triad:
\begin{itemize}
    \item ${ \hat{n}}$ in the direction of the test particle's orbital angular momentum.
    \item ${ \hat{u}}$ in the direction of the periapse of the orbit.
    \item ${ \hat{v}} = { \hat{n}} \times { \hat{u}}$.
\end{itemize}
The frame is of course fixed for unperturbed particle motion which takes place in the (${ \hat{u}},{ \hat{v}})$-plane, and osculates with the orbit in the presence of perturbations which are here assumed small in comparison to the pull of the central body. Bound Keplerian orbital motion is given by: 
\begin{equation}
    r=\frac{a(1-e^2)}{1+e\cos(f)}, \hspace{1cm} \vec r=r (\cos(f) \,{ \hat{u}} +\sin(f) \,{ \hat{v}})
\end{equation}
where $a$ and $e$ are the particle's semi-major axis and eccentricity, $r$ the orbital radius, and $\nu$ the true anomaly relative to ${ \hat{u}}$. 

Secular dynamical evolution  is then captured by time averaging any Hamiltonian contributions over a period $P= 2\pi/\sqrt{GM/a^3}$, on the particle's osculating Keplerian orbit. In so doing, it is useful to keep in mind the following differential relations
\begin{equation}
\frac{dt}{P}=\frac{rdE}{2\pi a}=\frac{r^2d f}{2\pi a b}=\frac{dM}{2\pi},
\end{equation}
where $b = a \sqrt{1-e^2}$ is the pericenter distance, \textit{E} the eccentric anomaly, and \textit{M} the mean anomaly. 

Similar expressions are used when further averaging over the mean anomaly of the outer perturber assumed on an eccentric Keplerian orbit with position vector $\vec{r}^\prime$, semi-major axis and eccentricity $a_{o}$, and $e_{o}$ respectively, and reference triad: $\, {\hat{n}_{o}}$, \, ${\hat{ u}_{o}}$, \, ${\hat{v}_{o}}$.

Secular test particle dynamics is then best parametrized with the normalized angular momentum vector, $\vec j =\sqrt{1-e^2} {\hat{n}}$ and the Lenz vector $\vec e  = e {\hat{u}}$, thus avoiding singularities of Keplerian orbital elements at zero eccentricity and/or inclination. Recovering expressions for orbit-averaged multipoles in terms of those vectorial elements is a straightforward though somewhat laborious exercise which is now well documented in various publications on hierarchical triples [e.g. \cite{Tremaine, correia2011tidal, hamers2020secular}]. With those and similar such works for reference, we simply quote the doubly averaged Hamiltonian associated with $H_p$:
\begin{align}
    H_S= -\frac{GM}{2a} +\frac{GM}{a}\Psi\label{basic_Ham}
\end{align}
with $\Psi = \Psi_p + \Psi_{quad} +\Psi_{oct}$, and
\begin{equation}\nonumber
    \Psi_p= \frac{\varepsilon_p}{4(1-e^2)^\frac{5}{2}}\bigg[1-e^2-3(\vec j\cdot\hat{n}_p)^2\bigg]
\end{equation}
\begin{equation}\nonumber
    \Psi_{quad}=\frac{3 \varepsilon_{o}}{8}\Big[\frac{1}{3}-2e^2+5e_n^2-j_n^2\Big]
\end{equation}
\begin{equation}\label{Hamiltonian_components_psis}
    \Psi_{oct}= \frac{15}{64}\varepsilon_{o}\varepsilon_{\otimes} \Big[e_u\big(8e^2-1-35e_n^2+5j_n^2\big)+10j_uj_ne_n\Big]
\end{equation}
with dimensionless parameters given by
\begin{equation}
 \varepsilon_p=\frac{Q_2^{\prime}}{a^2}\hspace{0.2cm};\hspace{0.2cm}
    \varepsilon_{o}=\frac{M_{o}a^3}{Ma_{o}^3(1-e_{o}^2)^{\frac{3}{2}}} 
\hspace{0.2cm};\hspace{0.2cm}
\varepsilon_{\otimes}=\frac{e_{o}}{1-e_{o}^2}\frac{a}{a_{o}}.
\end{equation}
Subscripts affected to $j$ and $e$ reflect bases vectors on which $\vec j$ and $\vec e$ are projected respectively.

Along with TTN, we remind the reader that in our model Hamiltonian: (1) the inner quadrupole is spatially fixed, a valid assumption when the precession rate is negligibly slow; (2) the particle of interest is massless; (3) the test particle is far enough from the inner perturber that its potential can be approximated as a quadrupole; (4) the outer perturber is sufficiently hierarchical (and eccentric) for its tide to be reasonably approximated in the adopted octupolar expansion, and at the same time sufficiently far from the inner perturber to neglect their mutual perturbation, and resulting precession (refer to the discussion section below for further elaborations on perturber precession); (5) finally, the obvious, namely that the perturbations are weak enough for the dynamics to be reasonably well captured in the secular orbit averaged framework.   

\subsection{Equations of Motion} \label{secular_EOM}
Secular dynamics of the test particle is dictated by $H_S$. This Hamiltonian is cyclic in the particle's mean anomaly, thus the momentum conjugate to that anomaly, $\sqrt{GMa}$, is conserved, and with it the semi-major axis of the orbit. The orbit's orientation and shape are then fully controlled by the evolution of $\vec e$ and $\vec j$. We follow \cite{allan1964long} and \cite{Tremaine} in writing the vectorial equations of motion generated by the orbit averaged Hamiltonian of Eq.\eqref{basic_Ham}:
\begin{align} \label{Mil_Je} \nonumber
     \frac{d \vec j}{d\tau}&=-\vec j\times\nabla_{\vec j} \Psi-\vec e\times\nabla_{\vec e}\Psi \\
     \frac{d \vec e}{d\tau}&=-\vec e\times\nabla_{\vec j} \Psi-\vec j\times\nabla_{\vec e}\Psi,
\end{align}
with $\tau=\sqrt{\frac{GM}{a^3}}t$. Expanding, we obtain

\begin{align}\nonumber
    \frac{d\vec j}{d\tau}&= \frac{3}{4}\varepsilon_o j_n\vec j\times\hat{n}_o-\frac{15}{4}\varepsilon_o e_n\vec e\times\hat{n}_o+\frac{3}{2}\varepsilon_p\frac{(\vec j\cdot\hat{n}_p)}{(1-e^2)^\frac{5}{2}}\vec j\times\hat{n}_p\\\nonumber
     &-\frac{75}{64}\varepsilon_o\varepsilon_\otimes\Bigg[\Big[2\big(e_uj_n+e_nj_u\big)\vec j
     +2\big(-7e_ne_u+j_uj_n\big)\vec e\Big]\times \hat{n}_o\\
     &+\Big[2e_nj_n\vec j+\big(-7e_n^2+j_n^2+\frac{8}{5}e^2 -\frac{1}{5}\big)\vec e
     \Big]\times\hat{u}_o\Bigg],
     \label{eomj}
\end{align}  

\begin{align}\label{eome}\nonumber
    \frac{d\vec e}{d\tau}&= \frac{3}{4}\varepsilon_o j_n\vec e\times\hat{n}_o-\frac{15}{4}\varepsilon_o e_n\vec j\times\hat{n}_o+\frac{3}{2}\varepsilon_p\frac{(\vec j\cdot\hat{n}_p)}{(1-e^2)^\frac{5}{2}
}\vec e\times\hat{n}_p\\\nonumber
    &+\Big[\frac{3}{2}\varepsilon_o -\frac{3}{4}\varepsilon_p \frac{1-e^2-5(\vec j\cdot\hat{n}_p)^2}{(1-e^2)^\frac{7}{2}}\Big]\vec j\times\vec e\\\nonumber
     &-\frac{75}{64}\varepsilon_o\varepsilon_\otimes\Bigg[\Big[2\big(e_uj_n+e_nj_u\big)\vec e
     +2\big(-7e_ue_n+j_uj_n\big)\vec j\Big]\times \hat{n}_o\\
     &+\Big[2e_nj_n\vec e+\big(-7e_n^2+j_n^2+\frac{8}{5}e^2 -\frac{1}{5}\big)\vec j
     \Big]\times\hat{u}_o
     +\frac{16}{5}e_u\vec j\times\vec e\Bigg].
\end{align}  

We pause for a few remarks before getting down to business:
\begin{itemize}
     \item \textit{The Octupole:} Terms involving $\varepsilon_\otimes$ reflect the octupolar perturbation by the eccentric outer body. The eccentricity of that body was already present at the quadrupolar level, but only as a constant factor through an averaged contribution which is independent of the perturber's angular orientation. Here, that orientation is explicit through $\hat{u}_o$. 
    
     \item \textit{Symmetry Breaking:} With orientation dependent terms, the octupole breaks symmetries that were present at the quadrupolar level. In particular, the equations of motion are no longer invariant when flipping the eccentricity vector $\vec e\rightarrow-\vec e$. Further flipping  $\vec{j}\rightarrow-\vec{j}$ takes the equations from  $\frac{d\vec{j}}{d\tau}\rightarrow -\frac{d\vec{j}}{d\tau}$, $\frac{d\vec{e}}{d\tau}\rightarrow -\frac{d\vec{e}}{d\tau}$, and one can restore invariance through time reversal, $\tau\rightarrow -\tau$. Orbit-wise, this is equivalent to taking $i\rightarrow\pi-i, \omega \rightarrow \pi -\omega, \Omega\rightarrow\Omega+\pi$.
    
    \item \textit{The Principal Plane:} Following TTN, we define the plane whose normal is oriented along $\hat{n}_p\times\hat{n}_o$ as the principal plane, and we denote by $\Theta$ the angle between those two vectors. Classically, this angle measures the obliquity of an oblate planet, but it can be generalized to the mutual inclination between the perturbations whatever the system is. Invariance under  $\hat{n}_o\rightarrow-\hat{n}_o$ implies that for a full parametric study, we can restrict $\Theta$ to the range $[0,\pi/2]$ instead of $[0,\pi]$.

    \item \textit{Equilibria:} TTN distinguished between \textit{coplanar-coplanar} and \textit{coplanar-orthogonal} equilibria. In the former, both $\vec j$ and $\vec e$  lie in the principal plane, while in the latter, one of the vectors is in the plane and the other orthogonal to it. In our case, and with the added complexity of the perturber's orientation, it was already challenging enough to study the \textit{coplanar-coplanar} configuration in all its glory, so we left other potential configurations for investigations to follow. Furthermore, we solve for \textit{coplanar-coplanar} in the simplest configuration where $\hat{n}_o, \hat{u}_o$ and $\hat{n}_p$ are in the same plane. Concerns about this condition will, we hope, be assuaged in the discussion below. 
    
    \item \textit{Stability:} When we speak of the stability of relative/Laplace equilibria of the dynamical system above, we are mainly referring to linear stability which is assessed by considering linearized dynamics around equilibrium angular momentum and eccentricity vectors. The procedure is straight-forward and will not be spelled out explicitly here. For the circular equilibria of the quadrupolar limit, TTN were able to differentiate between stability to perturbations in eccentricity then angular momentum, making use of the decoupling of the linearized equations. It was thus possible for them to derive elegant analytical expressions for the eigenvalues. With the octupole at play, the decoupling is no longer feasible, and we resort to solving for the eigenvalues of the relevant $6\times6$ matrix numerically. We tested our linear stability toolbox by confronting our generalized framework with TTN's results, then confirming linear stability results with direct integration of the full equations of motion in the neighborhood of equilibria. 
    \end{itemize}

\section{Whither the Laplace Surface?}

For equilibria of interest to us here, those fulfilling the \textit{coplanar-coplanar} condition described above, all vectors $\hat{n}_p,\hat{n}_o,$ $\hat{u}_o$, $\vec j$, and $\vec e$ are assumed to lie in the same principal plane. In this configuration, the nodes of the planes of all three players are aligned: the mutually inclined inner and outer perturber, and the test particle in between. Plus, the argument of the apse of the outer perturber is assumed frozen at 90 degrees from the ascending node on the plane of the inner perturber.  In this case, equilibrium conditions for a spatially frozen test particle orbit are deduced from Eqs.\eqref{eomj} and \eqref{eome} as they reduce to two scalar equations for two unknowns, the orbital eccentricity $e$ and the inclination angle between $\vec j$ and $\hat{n}_p$, which we denote by $\phi$. They are given by:

\begin{align}
\label{1st_eq_cond}
\nonumber
0=&\frac{3}{4}\varepsilon_p(1-e^2)^{-\frac{3}{2}}\sin(2\phi)-\frac{3}{8}\varepsilon_o(1+4e^2)\sin2\Delta\phi\\\nonumber
&\mp\frac{75}{64}\varepsilon_o\varepsilon_\otimes e\Bigg[-\frac{7}{2}(1+e^2)\cos\Delta\phi\sin2\Delta\phi\\
&+(2+5e^2)\sin^3\Delta\phi+\frac{1}{5}(1-8e^2)\sin\Delta\phi\Bigg]
\end{align}
\begin{align}
\label{2nd_eq_cond}
\nonumber
0&=\frac{3}{4}\varepsilon_p\frac{e}{(1-e^2)}(1-3\cos^2\phi)-\frac{3}{4}\varepsilon_o e\sqrt{1-e^2}(1-4\sin^2\Delta\phi)\\\nonumber
&\mp\frac{75}{64}\varepsilon_o\varepsilon_\otimes(1-e^2)^\frac{1}{2}\Bigg[(3e^2-1)\cos^3\Delta\phi\\
&+(\frac{1}{5}-\frac{24}{5}e^2)\cos\Delta\phi
+(1+\frac{15}{2}e^2)\sin\Delta\phi\sin2\Delta\phi\Bigg]
\end{align}
where $\Delta\phi=\Theta-\phi$. We distinguish between equilibria which are aligned and those which are anti-aligned with the external perturber's periapse. The two differ by the sign of the octupole terms, with the upper sign delivering the aligned configuration.

\subsection{Destroying the  \textit{Classical Laplace Surface}: The Octupole at Work}
Considering circular equilibria, ones with $e=0$, octupolar terms vanish from the angular momentum equilibrium condition \eqref{1st_eq_cond} leaving quadrupolar terms, while the opposite happens in the Lenz vector equation \eqref{2nd_eq_cond}:
\begin{subequations}
\begin{align}
0&=2\varepsilon_p\sin(2\phi)-\varepsilon_o\sin2\Delta\phi \label{quadrupole_circular}\\
0&=\varepsilon_o\varepsilon_\otimes \cos\Delta\phi\big(-3\cos^2\Delta\phi+\frac{11}{5}\big).\label{octupole_circular}
\end{align}
\end{subequations}
Solutions to those equations remain quite rich and are succinctly captured in Fig.\ref{circular_equilibria}. In what follows, we highlight key features:
\begin{itemize}
    \item In the quadrupolar limit studied by TTN (Eq.\ref{quadrupole_circular}), circular equilibria fall in two families: \textit{i)} a stable family that runs between $\phi=0$ as $a\rightarrow 0 $ to $\phi\rightarrow \Theta$ as $a\rightarrow \infty$ forming the  \textit{Classical Laplace Surface.} This transition between perturbation planes occurs around the Laplace radius given by
    \begin{align}
    r_L^5= Q_2^\prime  a_o^3 (1-e_o^2)^{3/2}\frac{M}{M_o} = a^5 \frac{\varepsilon_p}{\varepsilon_o}.
    \label{laplacesurfaceeq}
\end{align}
\begin{figure}
    \centering
    \includegraphics[width=\columnwidth]{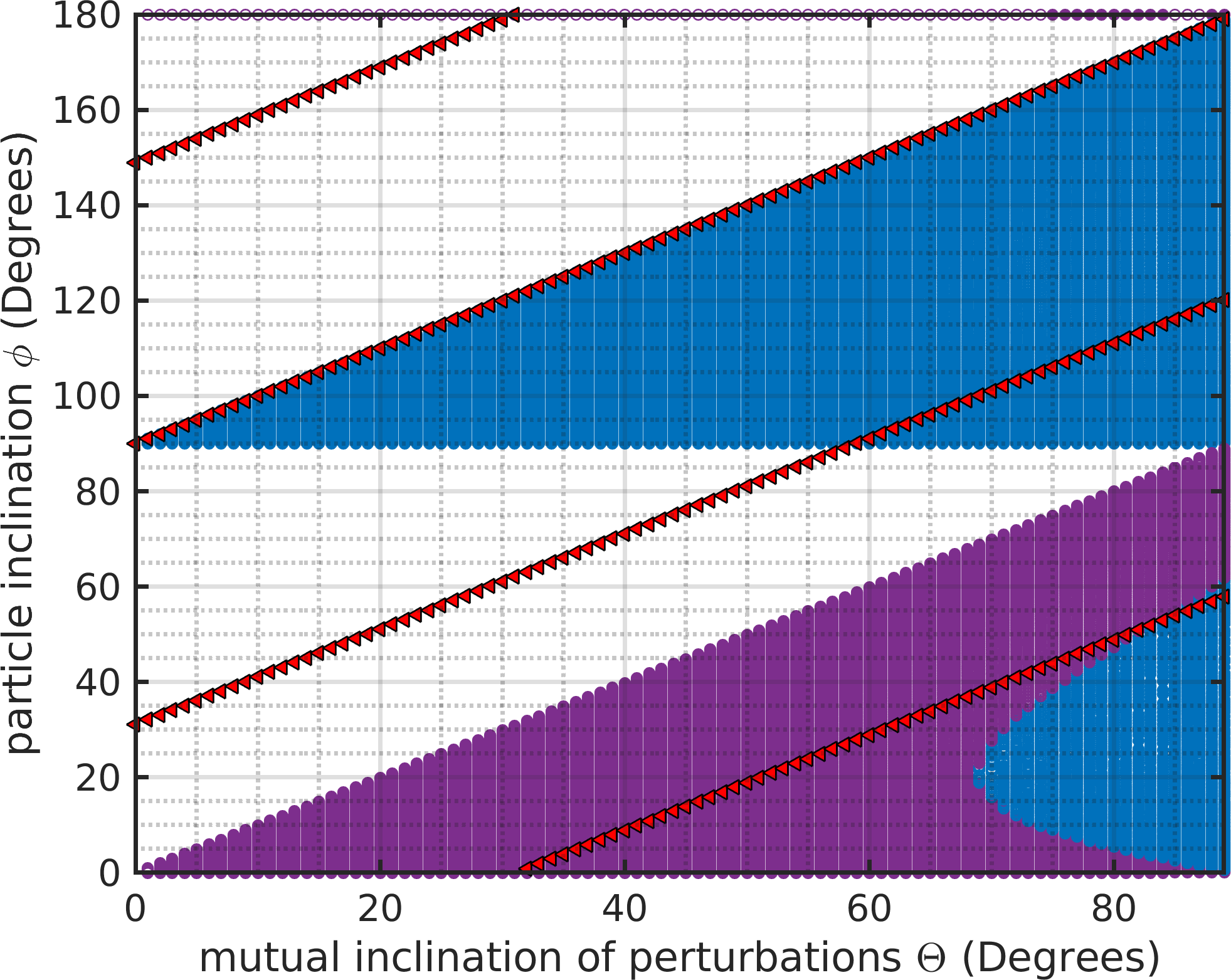}
   
    \includegraphics[width=\columnwidth]{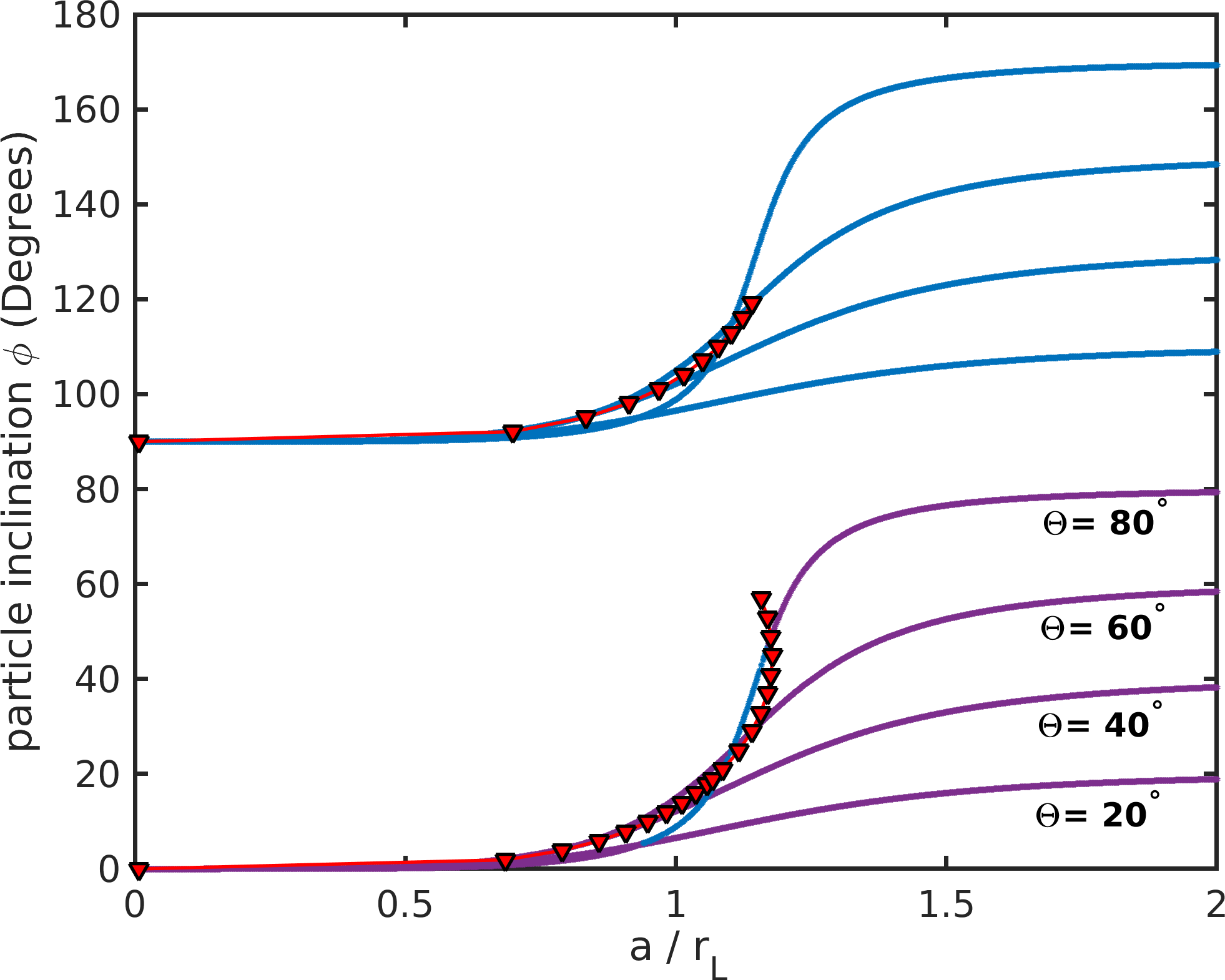}
    \caption{Inclinations of circular coplanar-coplanar Laplace equilibria. \textit{ Top}: Equilibrium inclination as a function of the mutual inclination between the inner quadrupole and the outer eccentric perturber ($e_o=0.65)$. Purple (stable) and blue (unstable) orbits satisfy the relative equilibrium conditions in the quadrupolar limit  (Eq.\ref{quadrupole_circular}). Designated by red triangles, on top of the quadrupolar solutions, are pairs that satisfy the condition introduced by the octupole (Eq.\ref{octupole_circular}). Pairs overlapping with the quadrupolar shaded regions are the survivors of the classical {\it Laplace Surface} upon the Octupolar addition. \textit{Bottom:} Inclinations of circular equilibria, now mapped as a function of semi-major axis, for a range of mutual inclinations ($\Theta= 20,40,60,80^\circ$).}  
    \label{circular_equilibria}
\end{figure}
TTN showed how circular Laplace equilibria on the \textit{Classical Laplace Surface} go unstable over a range of inclination and semi-major axis around the Laplace radius. This occurs when $\Theta$ exceeds a critical value of $68.875^\circ$. In the top panel of Fig.\ref{circular_equilibria}, we reproduce in purple TTN's stable equilibria, in the range $\phi=0\dots {\phi}_{o}$, enclosing the blue zone of unstable equilibria. \\   

\textit{ii)} Also shown in blue are TTN's unstable retrograde equilibria, running between $\phi=\pi/2$ as $a\rightarrow 0 $ to $\phi\rightarrow \Theta+\pi/2$ as $a\rightarrow \infty.$ In the second panel of Fig.\ref{circular_equilibria}, TTN's prograde and retrograde families are shown as a function of semi-major axis, for four different mutual inclinations between the perturbations ($\Theta=20,40,60,80^\circ$). 
   \item  When we account for the eccentricity of the outer perturber through the octupole terms, classical Laplace equilibria are further constrained by the additional condition in Eq.\eqref{octupole_circular} which imposes: 
\begin{equation}
\label{solutions_octupole_cirular}
    \cos\Delta\phi=0 \hspace{1cm } \text{or} \hspace{1cm } \cos\Delta\phi=\pm\sqrt{\frac{11}{15}}.
\end{equation}
Orbits satisfying this additional condition are displayed in red together with the quadrupolar solution in Fig.\ref{circular_equilibria}. With the octupole in action, surviving circular equilibria are given by those red dots which overlap with the shaded regions. As shown in the second panel of Fig.\ref{circular_equilibria}, classical families of equilibria now collapse into discrete equilibrium orbits, each corresponding to a distinct mutual inclination. Consequently, the warped \textit{Laplace Surface}, traditionally running over the full range of semi-major axis, is now destroyed and replaced by a distinguished family of circular orbits. It is interesting to note that surviving prograde circular equilibria would cease to exist beyond $a=1.17 r_L$, namely when entering the eccentric perturber dominated regime.  We also note that the structure of surviving relative equilibria is independent of the eccentricity of the outer perturber.

\item Though the destruction of structure requiring symmetry is perhaps not surprising, that circular equilibria survive this eccentric perturbation surely is. Those survivors carry the same stability signature as the quadrupolar regions they fall on. Even more curious perhaps is the surviving family with $\Delta\phi= 31.09^\circ$: a family of circular orbits which, independent of  the eccentricity of the outer perturber, are fixed in space with a constant tilt to the angle of mutual inclination between the perturbations. 
\end{itemize}

\subsection{The Emergence of the \textit{Eccentric Laplace Surface}}\label{ecc.lapl.surf}

Having explored what remains of the classical \textit{Laplace surface}, we now characterize the structure that replaces it by mapping the full set of equilibria in the coplanar-coplanar configuration,  without any prior constraint on eccentricity of the test particle. When it comes to model systems, one can envisage a satellite (the test particle) orbiting an oblate planet that revolves on an eccentric orbit around a star; or a multi-planetary system where the inner quadrupole is provided by an oblate star and/or coplanar inner planets, while the outer octupole is provided by an eccentric inclined distant Jupiter or a stellar binary companion. We proceed with the latter scenario, and carry out the exercise for a system with an inner stellar binary component of mass $1.5 M_\odot$ hosting a hot Jupiter of mass $0.5 M_J$  situated on a circular orbit at  $0.1$ AU, with a less massive ($1 M_\odot$) outer stellar binary companion revolving on an eccentric orbit with $a_o= 180 $AU and $e_o=0.95$. We use the test particle approximation for a hypothetical planet in between the mutually inclined perturbations.

\begin{figure}
    \centering
    \includegraphics[width=\columnwidth]{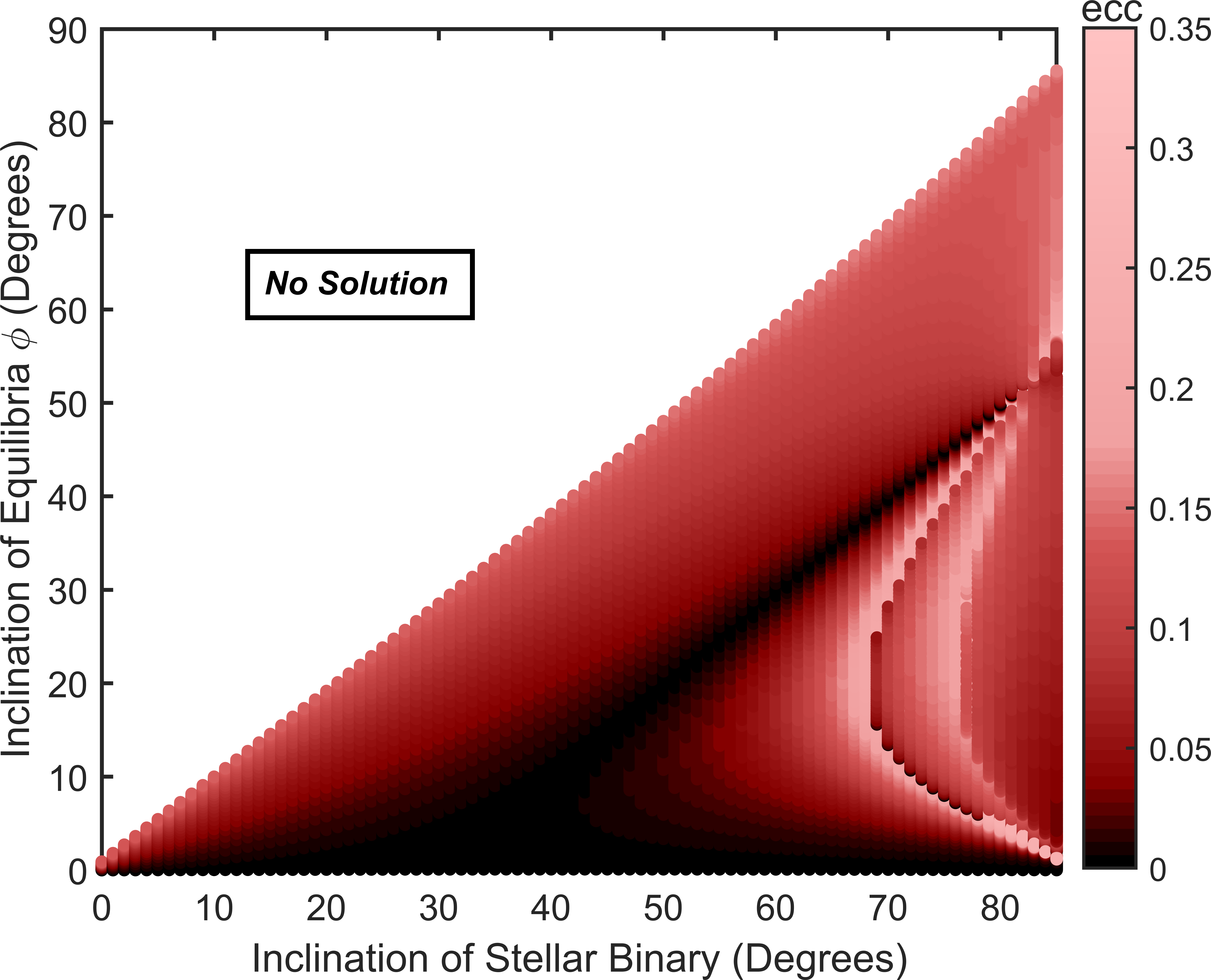}
    \caption{Inclinations of eccentric coplanar-coplanar equilibria as a function of the mutual inclination between the perturbations, with Octupole included. The studied system is the multiplanetary system of section \ref{ecc.lapl.surf}.  This figure is to be compared with the top panel of Fig.\ref{circular_equilibria}, corresponding to circular prograde equilibria. Similar to the quadrupolar limit, and for each mutual inclination between the perturbations, a family of equilibria starts in the plane of the inner quadrupole at small semi-major axis, then transitions to the plane of the outer eccentric perturber. However, equilibria are now eccentric due to the outer octupole, and their eccentricity increases as a function of the semi-major axis, thus creating the \textit{Eccentric Laplace Surface}. The color gradient codes for eccentricity, with darkest regions representing its smallest values, though not exactly zero.} 
    \label{E-LS}
\end{figure}

\begin{figure}
    \includegraphics[width=\columnwidth]{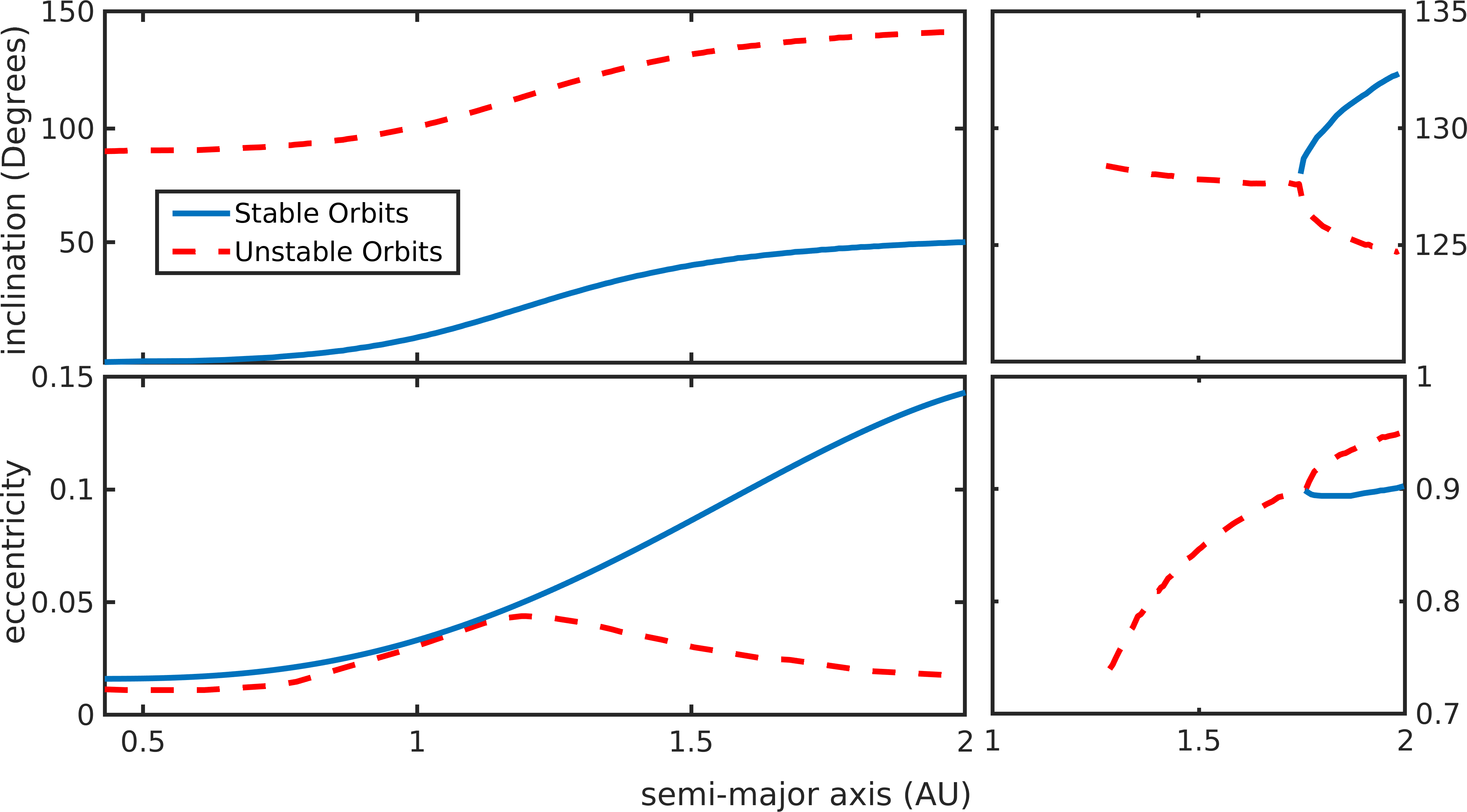}
    \caption{A sample of equilibria sectioned Fig.\ref{E-LS} and mapped into $(a,e)$ and $(a,i)$-space. The mutual inclination between the stellar companions is $50^\circ$. Compared to the quadrupolar limit, the stable Laplace Surface, along with its accompanying unstable family are shown in the panels on the left, preserving the inclination behavior but with eccentric stationary orbits. A highly eccentric retrograde bifurcation is evident in the panels on the right.  } 
    \label{multipoles_IB50}
\end{figure}

 The remnant circular equilibria of Fig.\ref{circular_equilibria} can then be situated within a continuum of eccentric and inclined equilibria which is mapped in Fig.\ref{E-LS}, then cross-sectioned in Fig.\ref{multipoles_IB50}. As apparent in Fig.\ref{E-LS} [which incidentally is limited to prograde orbits], eccentric equilibria display the same warped surface structure, interpolating between inner and outer planes, as they shape the  \textit{Eccentric Laplace Surface}. Equilibrium eccentricities increase with distance from the inner host as expected, reaching values around $e\approx0.2$ as they transition to the octupole dominated regime. Near-circular obits occupy two distinct regions: $\mathcal{R}_1$) towards the bottom of the figure, i.e. orbits dominated by the inner quadrupole ; $\mathcal{R}_2$) around the surviving family of circular equilibria with $\Delta\phi= 31.09^\circ$. The eccentric equilibria of TTN, those bifurcating from circular equilibria beyond $68.875^\circ$ are now part of a continuum of eccentric equilibria over the full range of mutual inclinations. 
 
For a perhaps more vivid appreciation of eccentric-inclined Laplace equilibria which emerge in the presence of an eccentric outer perturber, we map in Fig.\ref{multipoles_IB50} equilibrium families over a range of semi-major axes (straddling the Laplace transition), and a mutual inclination $\Theta=50^\circ$ between perturbers. The \textit{Eccentric Laplace Surface} is captured by the blue family of the left panels, a stable family of equilibria which transitions in inclination between the inner and outer planes (top left) as it increases in eccentricity (bottom left). In those same panels, we show how the  retrograde surface is maintained with highly inclined unstable equilibria of relatively small eccentricity. In the right panels of Fig.\ref{multipoles_IB50}, we isolate for clarity equilibrium families undergoing bifurcations of much higher eccentricity and inclination, into stable and unstable branches: much more to say about those below, as we consider implications for the shepherding of TNOs!
 
In sum, and when compared to TTN, the coplanar-coplanar skeleton of eccentric equilibria presented here reveals new and significant features, and this is both in the prograde \textit{Eccentric Laplace Surface} and the highly eccentric retrograde bifurcations. Prograde eccentric families in the quadrupolar limit of TTN are bound to two regions in $\Theta-\phi$ space: The region of circular equilibrium instability for $\Theta>68^\circ.875$ with $0<\phi<54^\circ.7$, and another region where $\Theta>54^\circ.7$ with $54^\circ.7<\phi<90^\circ$. Here we obtain them for the full range of mutual inclinations.

 We note that we are solving for equilibria over a carefully selected range of test particle semi-major axis $0.025 < \frac{a}{a_o}\frac{1}{1-e_o^2} < 0.12$. This choice is dictated by two requirements of the multipolar secular formalism: (1) avoiding close encounters between the test particle and its perturbers on one hand, (2) and making sure to guarantee the convergence of the multipolar expansion on the other. In this work, we have very little to say on potential departures from the averaging limit, but we do consider a way around the divergence of multipolar expansion, which yields an efficient means of exploring the full range of available secular equilibria. 

%% file: P9.tex
\section{FROM MULTIPOLES TO HARMONICS: DYNAMICS BEYOND NEPTUNE}
We extend our secular machinery to dynamical systems where the multipole expansion breaks down, providing a remarkably efficient remedy to the situation. We motivate our toolbox and associated results with a problem of current interest, namely the phase-space structure inhabited by Trans-Neptunian-Objects (TNOs) whose curious orbital architecture motivated the hypothesis of 9th planet in the outer parts of the solar system . Much has been written about this curious system \citep{BatyginBrown2016,holman2016observational,batygin2017dynamical,Li_P9,batygin2019planet} (including a contribution by one of the co-authors \citep{sefilian2019shepherding}) and it is not our objective here to review, discuss, defend or critique arguments or counter-arguments for one scenario or the other. Rather, we take the intellectual effort that has been exerted on the dynamics of this region of the solar system as a pretext to develop tools and insights (of the Laplace Surface variety), which we believe are of relevance to numerous other secular dynamical settings (from exo-planetary systems to black hole nuclei) where similar such mildly hierarchical structures obtain. The question for us reduces to the characterization of the secular orbital architecture of test particles, perturbed on the inside by the Giant Planets, and on the outside, by a putative 9th planet, a super-Earth, revolving on an eccentric and inclined orbit.

\begin{figure}
    \centering
    \includegraphics[width=0.9\columnwidth, height=7cm]{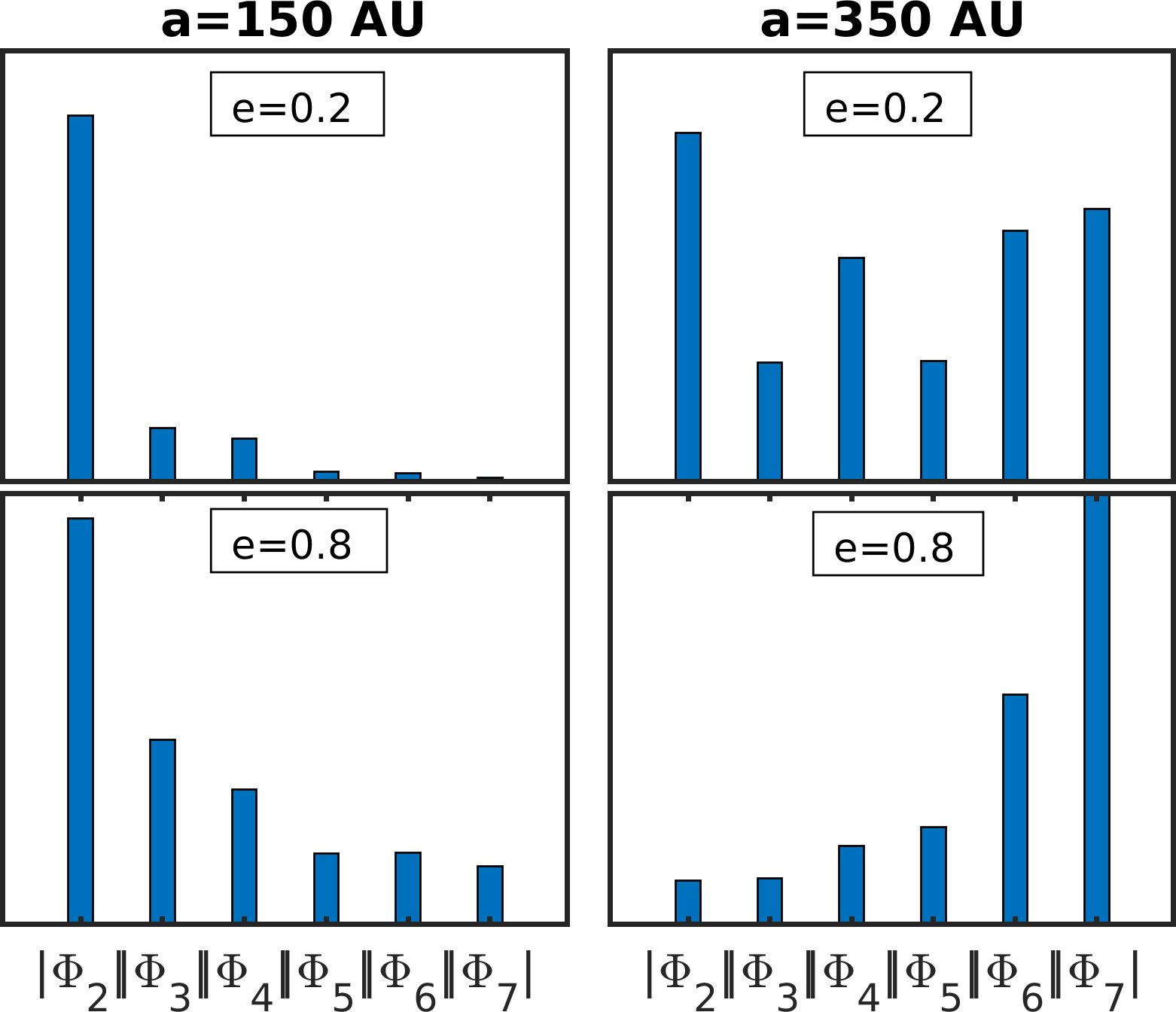}
     \caption{Numerical analysis of the increasing orders in the orbit averaged multipolar expansion of the disturbing potential of P9  up to 7th order  (normalized by the quadrupole $\Phi_2$). The potential is computed in the planar case for different orbital parameters of the test particle, with P9 at $a_9=700$ AU and $e_9=0.6$. The series converges for $a=150$ AU, for which the particle's apoapse falls within the orbits-crossing limit for any eccentricity. However, the series diverges at a higher semi-major axis ($a=400$ AU), for which the apoapse falls beyond the orbits-crossing limit for both presented eccentricities. }
      \label{sample_convergence}
\end{figure}

The orbital configuration of P9 is in a process of continued refinement \citep{batygin2019planet, fienga2020new}. Here, we adopt the orbital parameters first introduced in \citep{BatyginBrown2016}, namely $a_9=700$ AU, $e_9=0.6$, and a slightly inclined orbit with $i_9=10^\circ$. Considering this configuration (and updates that followed), together with the distribution of TNO semi-major axes, it is evident that the hierarchy of the hypothesized system is rather weak; a multipole expansion in the ratio of semi-major axes is expected to fail. Indeed, and as one gathers from multipole coefficients displayed in  Fig.\ref{sample_convergence}, the series converges when TNO apo-apse is less than 280 AU (i.e. the peri-apse of P9) and diverges beyond. Our conclusion is consistent with studies of the convergence of the direct part of the disturbing function \citep{migaszewski2008secular,migaszewski2009equilibria}, and can be succinctly expressed with:
\begin{align}
\nonumber \label{orbits_crossing_condition}
    a(1+e)&<700(1 - 0.6)  \mbox{ AU}  \\
    &< 280 \mbox{ AU}.
\end{align}
In the model binary system which we explored in section \ref{ecc.lapl.surf}, this limit is satisfied for the full range of semi-major axes and eccentricities considered. To overcome this hurdle in the P9 context, relatively recent works resorted to numerical averaging of the disturbing function \citep{beust2016orbital, saillenfest2017non}. The approach was adopted for coplanar perturbers, though one should, in principle, be able to extend it to fully spatial interactions. It was costly to undertake in the co-planar case, it is expected to be even more so for fully triaxial configurations. 
Here, we adopt a somewhat more brutal approach, spreading P9's mass over the corresponding Gaussian ring (the result of averaging over P9's mean anomaly), then resolving the gravitational potential of the resulting P9-ring into spherical harmonics. Further numerical averaging of the dominant harmonics over a test particle's mean anomaly yields the desired Hamiltonian controlling the secular dynamics TNOs over the full range of eccentricity and inclination, and the range of semi-major axis of interest.  The full procedure is spelled out in Appendix \ref{App_P9_potential}.

\begin{figure}
    \centering
    \includegraphics[width=\columnwidth, height=7cm]{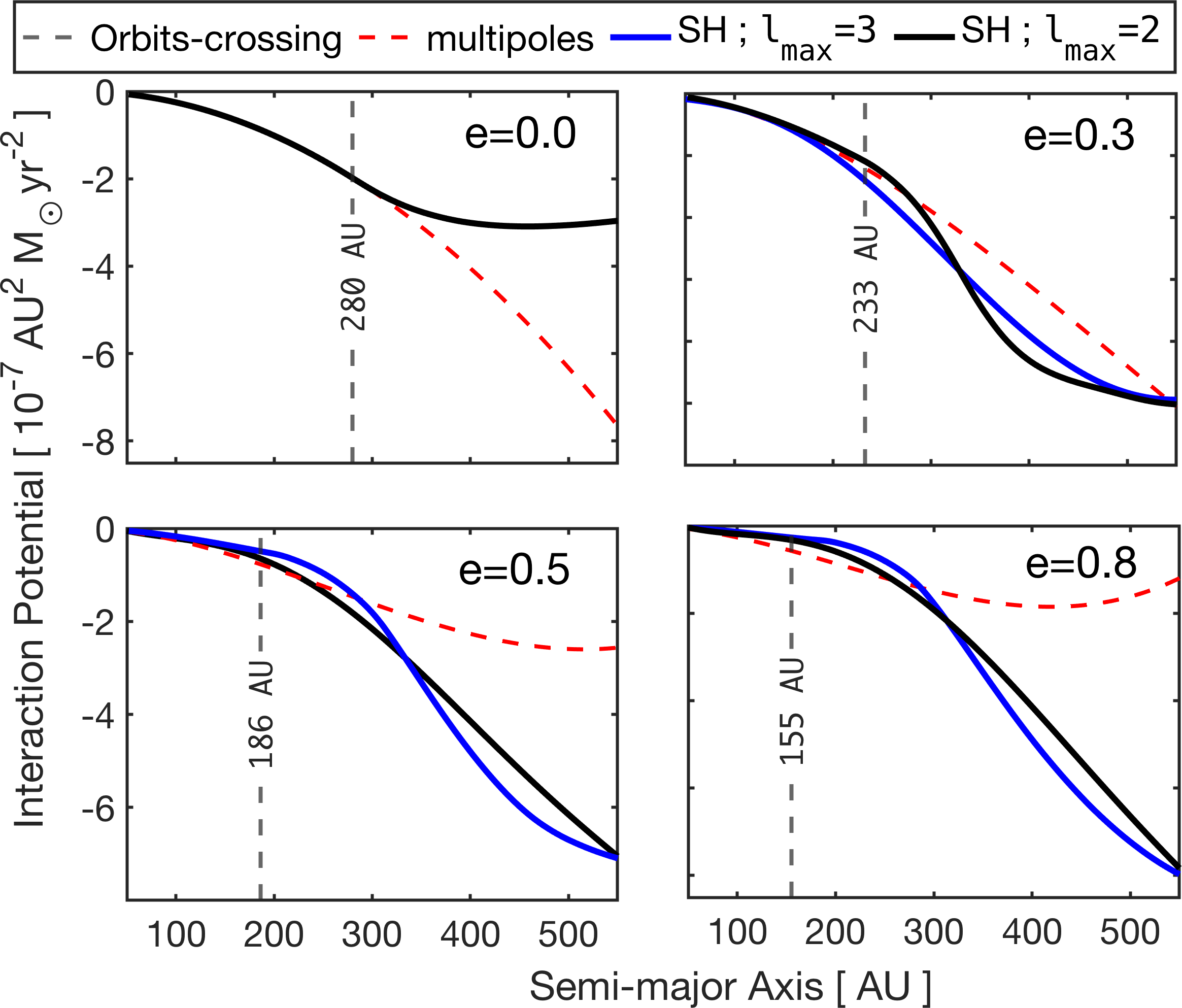}
   \caption{Comparison between P9 potentials generated by the orbit averaged multipoles expansion up to the octupole, and the orbit averaged numerical modes expanded in spherical harmonics. We distinguish in the latter between truncating at harmonic orders $l_{max}=2$ and $l_{max}=3$. The eccentricities are for a test particle evolving in the plane of P9, apsidally anti-aligned with it, and feeling its potential. For each eccentricity we identify the orbits-crossing limit defined by Eq.\eqref{orbits_crossing_condition} beyond which P9's ring contributes from within and without, and the multipoles expansion diverges. The two methods agree within this limit, with the lower order harmonics yielding a better agreement.}
    \label{comp_SH_MH}
\end{figure}

In Fig.\ref{comp_SH_MH}, we compare averaged harmonics thus constructed with the by now familiar octupolar series, noting reasonable agreement for semi-major axes which avoid the orbit crossing limit at any given eccentricity. The two approaches differ significantly, as they must, beyond that limit, with the multipolar series diverging, and harmonics yielding results that approach exact quadrature, as they decay in magnitude with increasing order (Fig.\ref{modes_fig}). The difference arises from the way each approach handles the mass distribution in the perturber's ring. For a given test particle, a sound mutltipole expansion assumes a point mass perturber which is either inside or outside that particle, and this on every point of its orbit. Thus for a TNO, if its ring falls totally within (without) the perturber's ring i.e. if the condition $r_{i}< r_{P9}$ ($r_{i}>r_{P9}$) is satisfied over its complete orbit, the multipoles expansion is reasonably adequate and compares favorably with the adopted expansion in spherical harmonics (Fig.\ref{comp_SH_MH}).

This condition can be easily verified in concentric circular orbits. However, with particles and perturber on eccentric orbits, a particle's orbit has to be always within the perturber's periapse, or always outside its apoapse, for the multipolar description to converge. For eccentric TNOs of interest, the mass distribution of P9 is at times within their orbit, at others outside. The proper expansion in this case is the so called Laplace expansion, with a switch in the ratio of radii reflecting the switch in hierarchy with respect to the central body. Orbit-averaged multipoles are unable to account for such a switch, whereas the orbit averaged harmonics capture it automatically, by recovering the potential of the ring as a whole. 

\subsection{Sanity Check: Co-planar Dynamics}
To test the validity of our numerically generated potential, we recover Laplace equilibria in a configuration where P9, the inner quadrupole, and test particles of interest reside in the same plane. This setting was explored with various approaches to orbital averaging \citep{BatyginBrown2016, beust2016orbital, saillenfest2017non, batygin2019planet}. In our case, the Hamiltonian  governing the particle's dynamics reduces to
\begin{equation}\label{Planar_Hamiltonian}
      \Bar{\Phi}_{\text{planar}}=  - \frac{\Gamma}{3 l_p^3} + \Bar{\Phi}_{\text{P9,planar}}
\end{equation}
where the first term corresponds to the inner quadrupole, with $\Gamma$ capturing the orbit averaged forcing of the giant planets (Eq.\ref{Gamma}), $l_p= |\vec j|= \sqrt{1-e^2}$, and  $\Bar{\Phi}_{\text{P9,planar}}$ is the planar restriction of the numerically computed 3D potential of P9 (Eq.\ref{p9_potential_harmonics}), which can be written as
\begin{align}
\nonumber 
  \Bar{\Phi}_{\text{P9,planar}}(a,e)&=\Upsilon_0(a,e) + \Upsilon_1(a,e) e \cos\Delta\varpi +\Upsilon_2(a,e) l_p^2  \\\nonumber 
  &+\Upsilon_5(a,e)e^2\cos^2\Delta\varpi+ \Upsilon_6(a,e)l_p^2 e\cos\Delta\varpi \\
  & + \Upsilon_{10}(a,e)e^3 \cos^3\Delta\varpi
\end{align}
where $\Delta\varpi$ is the particle's apsidal separation from the fixed apsidal orientation of P9, and the functions $\Upsilon_i(a,e)$ are defined in Eqs.\eqref{Phis_P9}.  

\begin{figure}
\centering
\includegraphics[width=\columnwidth]{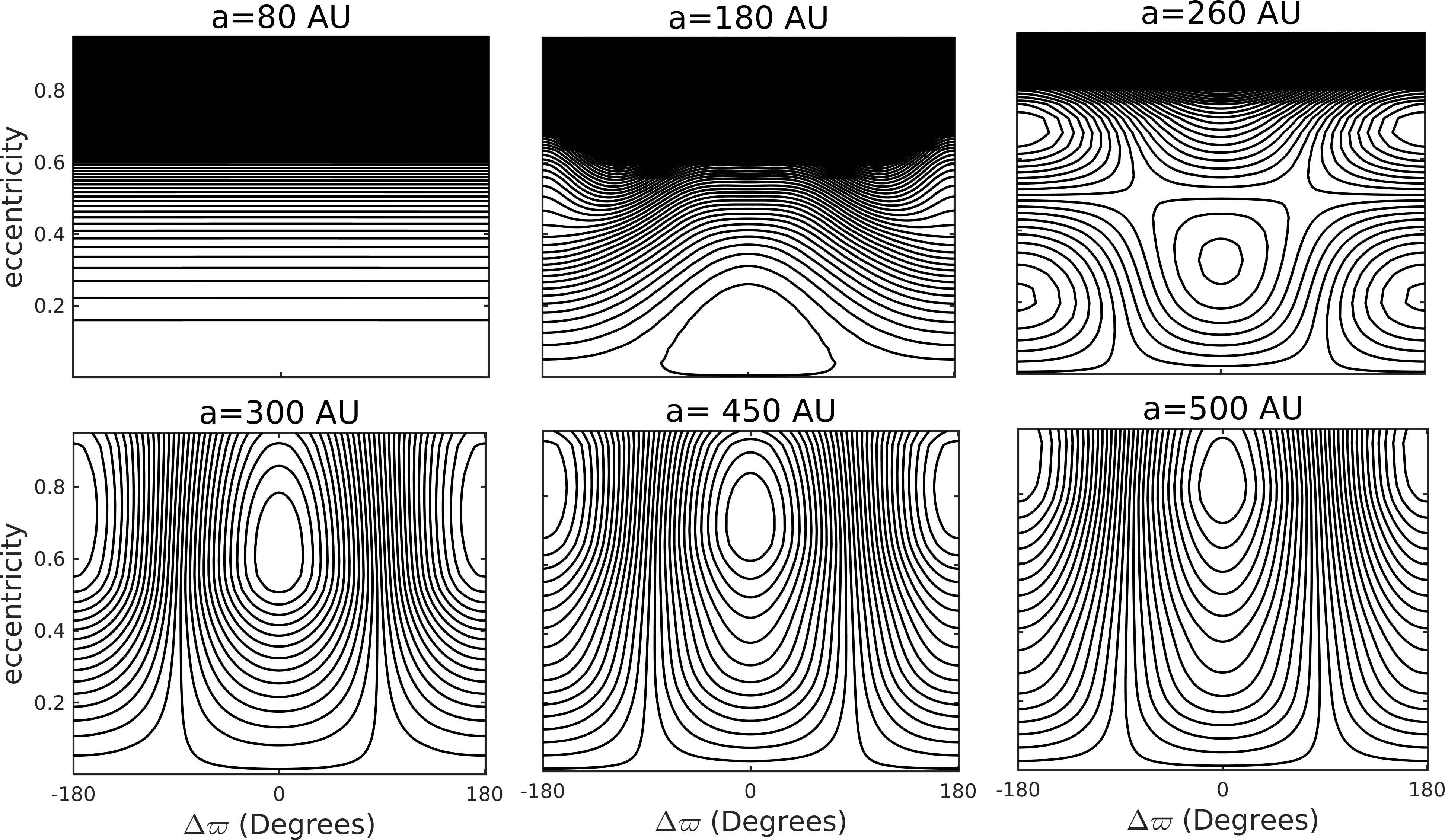}
   \caption{Phase portraits in $(\Delta\varpi,e)-$space corresponding to the planar Hamiltonian of Eq.\eqref{Planar_Hamiltonian} in which P9's potential is developed using the numerical harmonics toolbox. The chosen semi-major axis values cover the range over which the eTNOs are observed (see Table \ref{eTNOs_table}). For a relatively small semi-major axis of $a=100$ AU, the apsidal angle circulates. Beyond $a=150$ AU, TNOs can be captured in apsidally aligned or anti-aligned libration islands. }
    \label{Phase_Portraits}
\end{figure}

\begin{figure}
\centering
   \includegraphics[width=0.8\columnwidth]{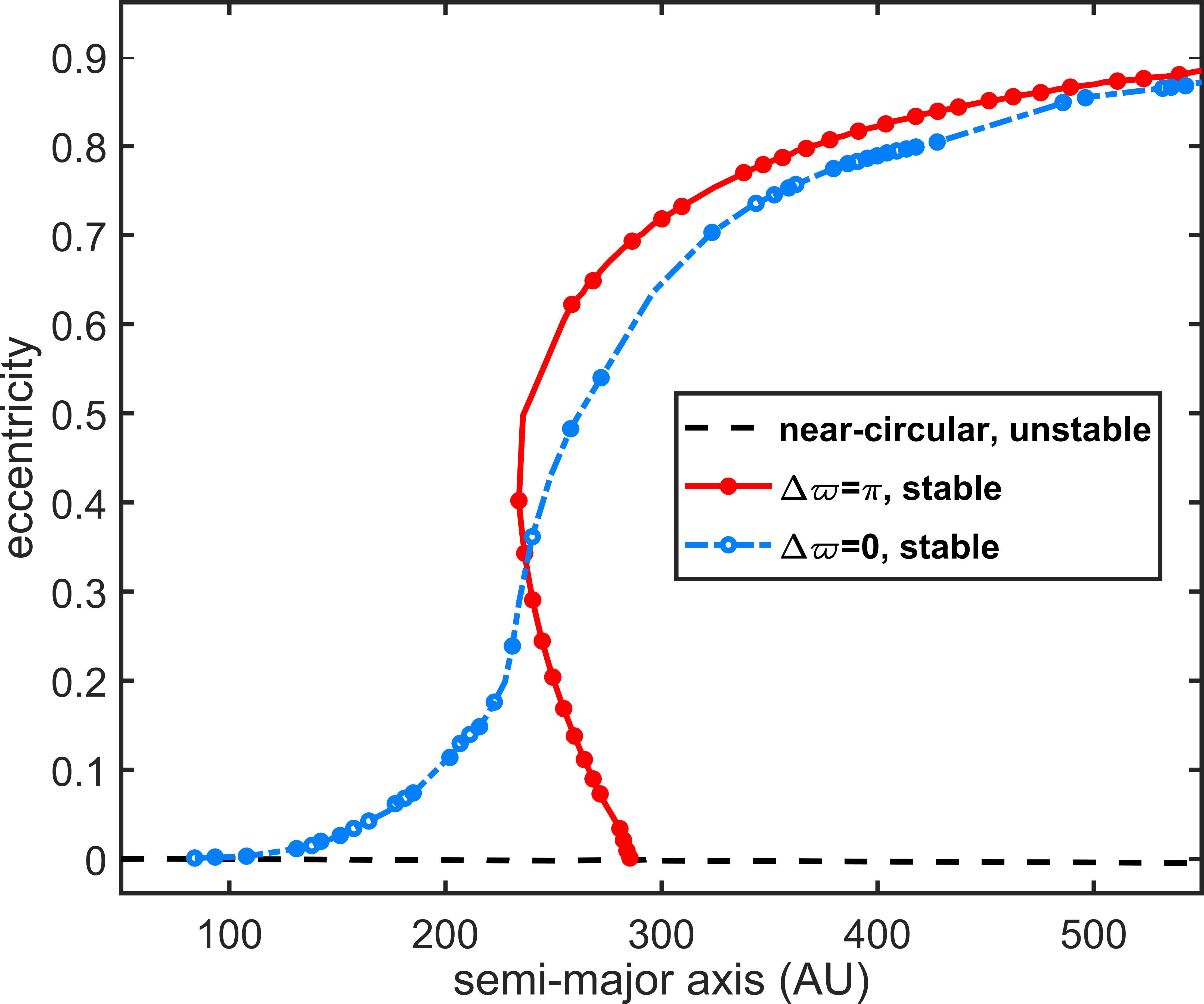}
   \caption{The eccentricity profile of stationary families of equilibria, both stable and unstable, apsidally aligned  ($\Delta\varpi=0)$ and anti-aligned ($\Delta\varpi=\pi)$ with the outer perturber, produced by solving the equation of motion \eqref{planar_EOM} over a range of semi-major axis $a$. The setting corresponds to the co-planar configuration where the giant planets, P9, and the test particles all live in the same plane. }
     \label{Planar_Equilibria}
\end{figure}

We capture phase portraits for a sequence of particle semi-major axes in Fig.\ref{Phase_Portraits}, then display Laplace equilibria for this co-planar configuration in Fig.\ref{Planar_Equilibria}. One family of stable (apsidally) aligned equilibria shows eccentricity growth with increasing semi-major axis. Another family of anti-aligned equilibria bifurcates with two stable branches beyond $a=234$ AU: one branch showing decreasing eccentricity with increasing $a$ until it vanishes for $a>290$ AU; the other growing in eccentricity with increasing $a$, surviving beyond $a= 290$ AU, reaching values as extreme as $e\approx 0.9$ by $a=550$ AU. Equilibria, their bifurcations, and associated phase-space structure are largely analogous to those presented by \cite{beust2016orbital} using numerical averaging of the full interaction potential. We further note agreement with trajectories of low inclination TNOs presented in \citep{Li_P9} (with slight differences likely resulting from differences in P9 parameters). 

In sum, using a rough and dirty shortcut to numerical averaging, we recovered key dynamical features in a weakly hierarchical problem, matching them to counterparts which were recovered with rather costly, nearly exact alternatives. At this stage, we hope we have given our reader enough confidence to follow us into the more treacherous triaxial landscape! 

\subsection{Moving on and Out: Fully Spatial Dynamics}
To explore life beyond the co-planar setting, we introduce a finite tilt to the orbit of P9.  We stick to the coplanar-coplanar equilibria of Section \ref{secular_EOM}, thereby reducing the three-dimensional vectorial equations of motion [Eqs. \ref{p9_EOM3d_dj} and \ref{p9_EOM3d_de}] into two coupled scalar equations for $(e,i)$, with three input parameters: the particle's semi-major axis $a$, its apsidal orientation $\Delta\varpi$, and the mutual inclination between the perturbers $\Theta$. For what follows, we shall pin our reference frame to the perturber's orbit, setting its inclination to zero, and its argument of perisapse to $\pi/2$. 

\begin{figure}
   \centering
    \includegraphics[width=0.9\columnwidth]{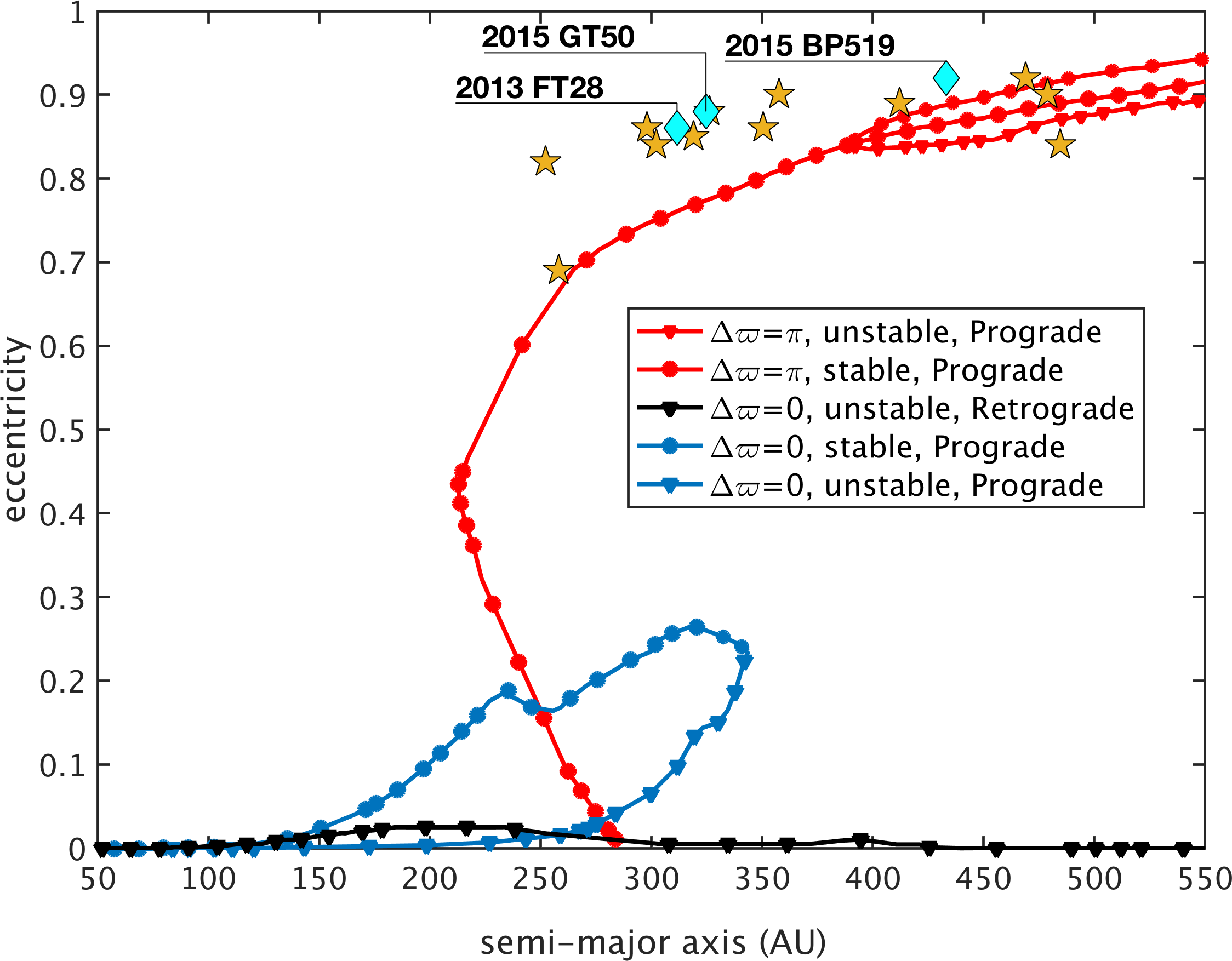}
    \includegraphics[width=0.9\columnwidth]{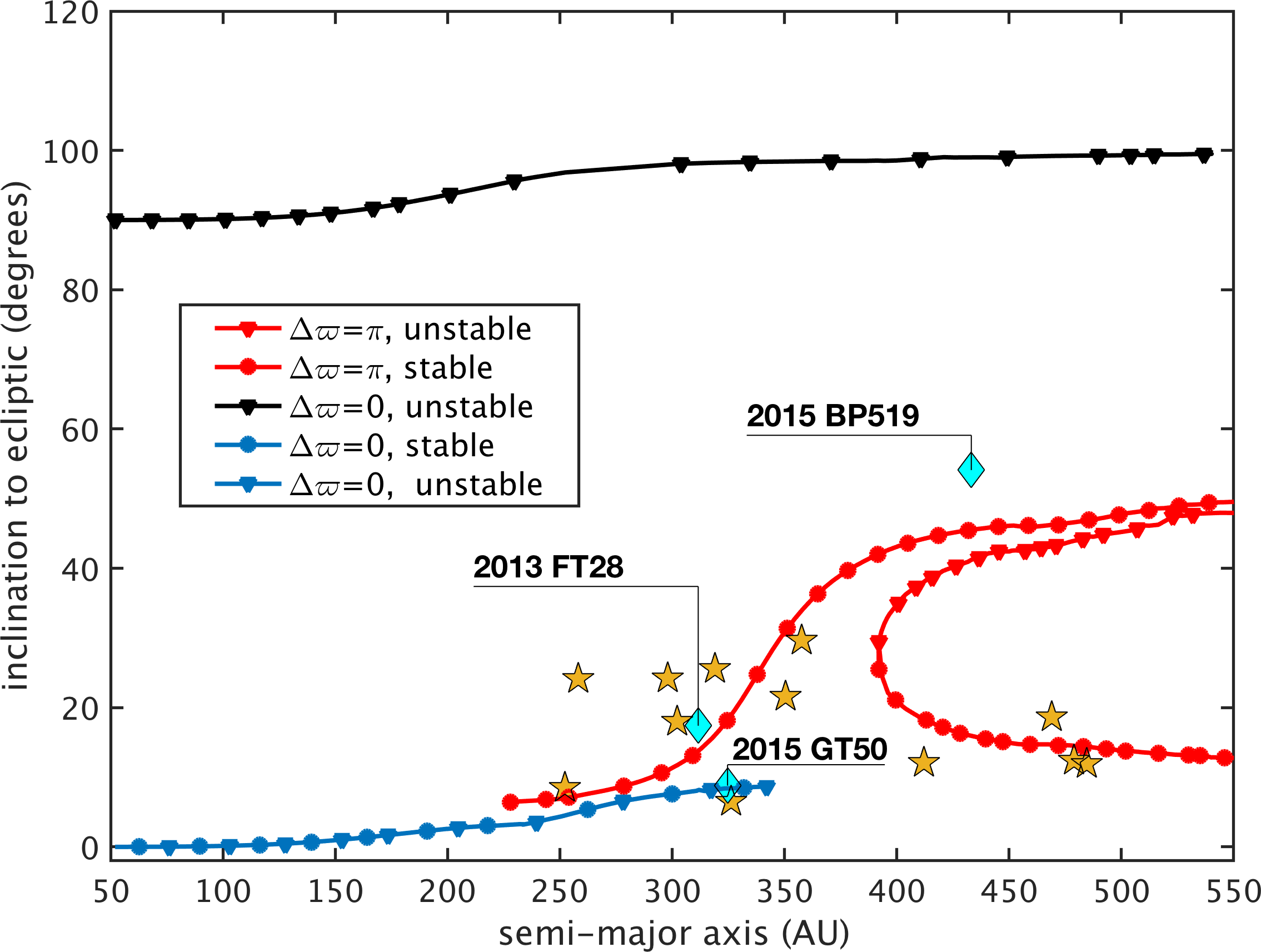}
   \caption{Equilibria of particles under the perturbation of the giant planets  and the quasi-numerically computed effect of an inclined P9 (with $\Theta=10^\circ)$. \textit{Top:} Eccentricity profile of equilibria as a function of $a$. \textit{Bottom:} Inclinations profile of equilibria as a function of $a$ (inclinations are transformed here to the plane of the inner planets). Equilibria apsidally aligned (anti-aligned) with P9 are plotted in blue (red). Families of stable (unstable) equilibria are plotted in circles (triangles). Plotted on top of the equilibrium profile are the observed eTNOs of Table \ref{eTNOs_table}.} 
   \label{3d_Equlibria}
\end{figure}

In Fig.\ref{3d_Equlibria}, we present a sample of equilibria in $(a,e)$ and $(a,i)$ space for $\Theta =10^\circ$. Pretty much as in the fully planar setting of Fig.\ref{Planar_Equilibria}, the stable, apsidally anti-aligned family persists when we tilt the perturber, though now starting beyond $a=213$ AU. It features an analogous eccentricity profile with semi-major axis, before two additional anti-aligned families bifurcate for $a > 391$ AU:  one stable and with larger eccentricity, the other unstable and with smaller eccentricity, than the central stable family. The persisting anti-aligned family now acquires inclination with increasing semi-major axis, reaching $\simeq 50^\circ$ by $500$ AU, at which point the equilibrium eccentricity is close to $0.9$. The bifurcating families emerge around an inclination $\simeq 25^\circ$, the unstable branch increasing in inclination towards $50^\circ$, while the inclination of the stable branch falls towards $\simeq 18^\circ$. 

The aligned family, on the other hand, bifurcates into two distinct branches in the $(a, e)-$plane, one stable and the other not, before disappearing altogether beyond $a>342$ AU. In the $(a,i)-$plane, those two branches lie on a Laplace-Surface of sorts which is stumped before hitting the outer plane. This is likely due to the inability of the giant planets to overcome the aligned particles orbital precession due to the strong couple exerted by P9 in this outer region. Upon further testing, we find that a quadrupole of strength $\Gamma^\prime = 2\Gamma$ suffices to maintain the highly eccentric aligned family, pretty much as in the planar setting, but laying it on a complete \textit{Eccentric Laplace Surface.}

One can further note the presence of the retrograde and unstable Laplace structure in Fig.\ref{3d_Equlibria}. This family starts in a polar configuration and turns increasingly retrograde  with distance, to eventually land around $\phi=\pi/2 + \Theta$, all the while maintaining a near circular shape (near and never exactly so). 

Overlaid over equilibrium profiles in Fig.\ref{3d_Equlibria} are the elements of clustered TNOs,  (Table \ref{eTNOs_table}), the majority of which are apsidally anti-aligned with P9, with eccentricities and inclinations following quite closely families of stable Laplace equilibria. This is not the case of "aligned" TNOs whose dynamics we discuss further below. We note that this equilibrium structure is robust to changes in $\Theta$ over a range of acceptable P9 inclinations: the eccentricity profile is maintained, with inclinations essentially following the mutual inclination. 

\begin{table}
\caption{Orbital parameters of observed  extreme TNOs of semi-major axis $250$AU$<a<650$AU and pericenter distance beyond Neptune. Data collected from the IAU Minor Planet Center in July 2020.} 
\centering 
\begin{tabular}{c c c c c c} 
\hline\hline 
TNO & $a $ & $e$ & $i$ & $\omega$ & $\Omega$ \\ [0.5ex]
    & $(AU)$  &  &$(^\circ)$  & $(^\circ)$   & $(^\circ)$\\  [0.5ex]
\hline 
2018 VM35 & 252.33 & 0.82 &  8.5 & 302.9 & 192.4 \\
2012 VP113 & 258.27 & 0.69 & 24.1 & 293.5 & 90.7 \\ 
2014 WB556 & 298.01 & 	0.86 & 24.2& 234.5& 114.8\\
2014 SR349 & 302.23 & 	0.84 & 17.9	 & 	340.9 & 34.8 \\
2013 FT28 & 311.61 & 0.86	 & 17.3	 &40.5 & 217.8	\\
2004 VN112 & 318.97 & 0.85 & 25.6 & 	326.8& 66.0  \\
2015 GT50 & 324.66 & 0.88 & 8.8 & 129.3& 46.1 \\
2013 SL102 & 326.18  & 0.88& 6.5 & 265.4 &  94.6 \\
2010 GB174& 350.59 & 0.86 & 21.6 & 	347.45 & 130.8   \\
2013 RF98 & 357.63 & 0.90 & 29.6 & 311.6 & 67.6 \\
2015 RX245 &411.98 &0.89 & 12.1 &65.1& 8.6 \\
2015 BP519 &433.17&0.92&54.1&348.2&135.0\\
2007 TG422& 468.98 &0.92 & 18.6& 285.6 & 112.9	  \\ 
2013 RA109&478.90&0.90&12.4&262.8&104.7\\
SEDNA & 484.52 &0.84 & 11.9& 311.5 &144.3	  \\ 
[1ex] 
\hline 
\end{tabular}
\label{eTNOs_table} 
\end{table}

\subsubsection{Poincar\'{e} sections locate Laplace in a sea of chaos}
To further examine dynamics in the neighborhood of Laplace equilibria, then of specific TNOs, we construct Poincar\'{e} sections on the ($\omega, e$)-plane with $\Omega=0$. Restricting to crossings with $\Dot{\Omega}<0$, we follow orbits governed by Eqs.\ref{p9_EOM3d_dj}-\ref{p9_EOM3d_de} and with initial conditions selected from an energy hypersurface. Producing sections over an interesting range of energies at various representative values of $a$, we highlight key features:

\begin{itemize}
    \item $a=60$  AU (Fig.\ref{ss-a60}): We start with a rather tame case at a relatively small semi-major axis where a TNO's dynamics is largely dominated by the inner quarupole, with near conservation of the vertical angular momentum $\sqrt{1-e^2}\cos i$. Dynamics on the section is fully regular, with a persistent libration zone around $\omega=\pi/2$, and a bifurcation at $\omega=-\pi/2$ with decreasing energy. Relative equilibria have inclinations $\approx 64^\circ$, and they correspond to periodic orbits, with finite $\Dot\Omega$, as opposed to $\Dot\Omega=0$ in the case of Laplace equilibria. The behavior is consistent with outer Kozai-Lidov dynamics associated with the inner quadrupole, dynamics which will get gradually encroached upon as we progress to larger semi-major axes \citep{saillenfest2017non}.

\begin{figure}
    \centering
     \begin{subfigure}[b]{0.45\columnwidth}\hfill
         \includegraphics[width=\columnwidth,height=3.7cm]{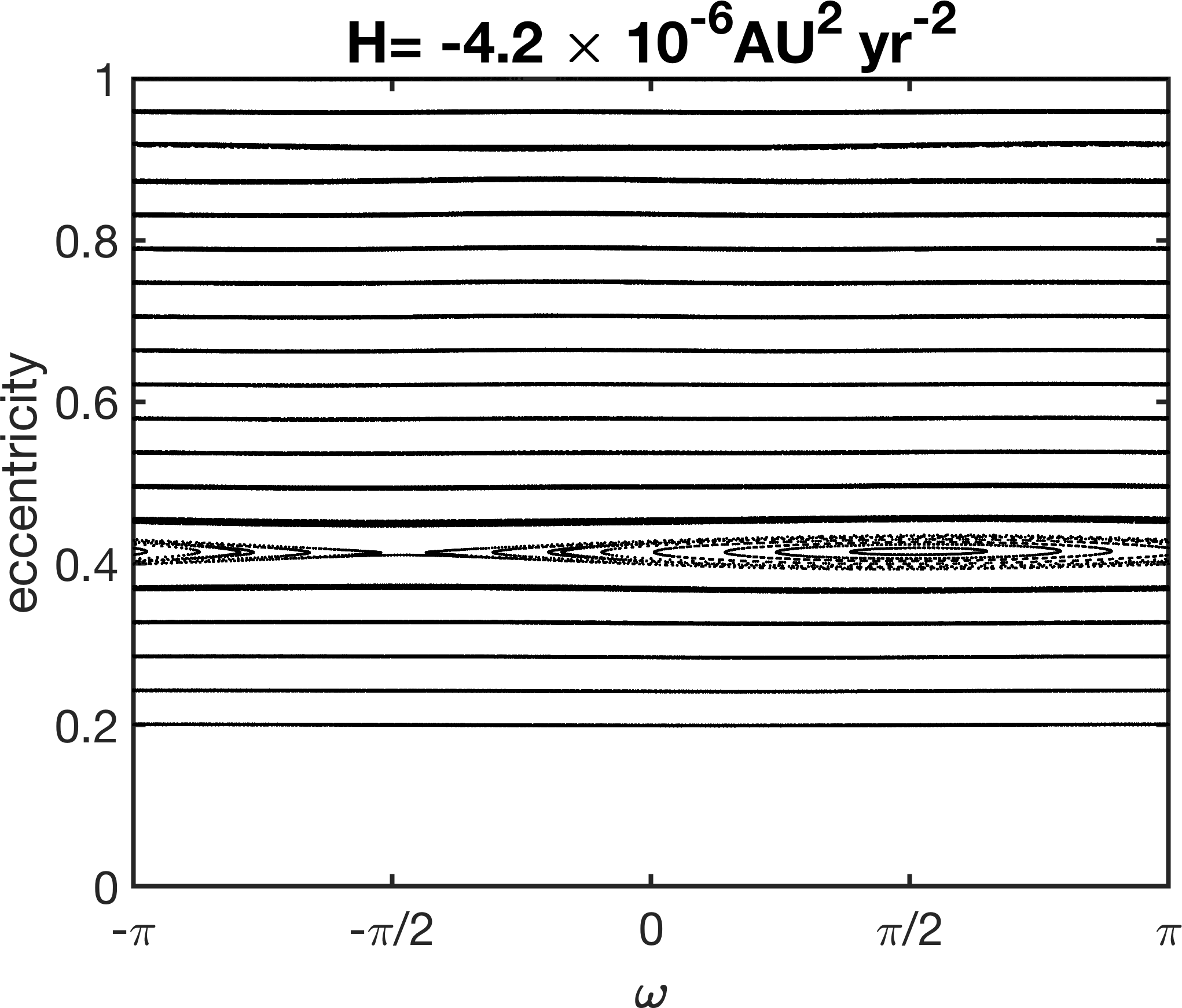}\hfill
    \end{subfigure}  
    \begin{subfigure}[b]{0.45\columnwidth}
         \includegraphics[width=\columnwidth,height=3.7cm]{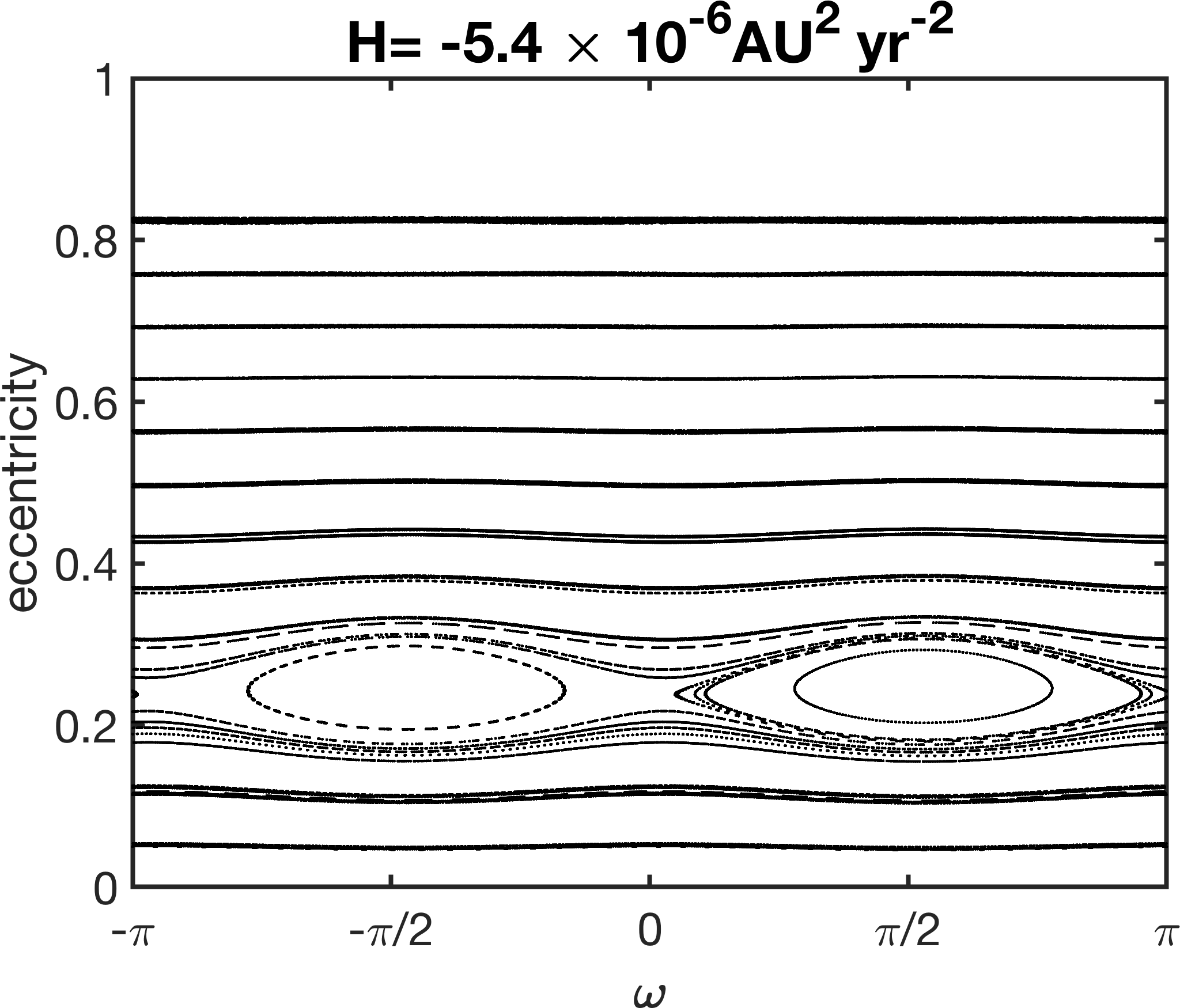}\hfill
    \end{subfigure}

    \caption{Poincar\'{e} sections for a test particle with $a=60 $AU, driven from the inside by the giant planets and from the outside by a hypothetical ninth planet. Initial conditions are selected on an energy hyper-surface, and trajectories sectioned in the $(\omega,e)$-plane for $\Omega=0$, and crossings with $\dot{\Omega} <0$. The left panel reveals fully regular motion over the whole range of eccentricity, with stable small amplitude librations in eccentricity around $e \approx 0.4$ and $\omega=\pi/2$, and an unstable fixed point around $\omega=-\pi/2$. In the right panel, we explore dynamics at a lower energy, with stable librations around $\pm \pi/2$, and the smaller eccentricity of $\approx 0.2$. Fixed points correspond to period orbits with inclination around $64^\circ$.}
    \label{ss-a60}
\end{figure}

\item  $a=258$  AU (Fig.\ref{ss-a258}): The interplay between the inner and external perturbers is now manifest, with chaos emerging around libration zones, and chains of resonant islands embedded within them. At $H= -5.3 \times 10^{-6} AU^2 yr^{-2}$, we follow dynamics with the secular energy of 2012 VP113. The aligned libration zone is encroached upon with chaotic trajectories, while the anti-aligned resonance persists. Below this libration zone, we plot in red the torus "associated" with this TNO [which is known to be stable \citep{batygin2019planet}]. We observe circulation from anti-aligned to aligned configurations [occurring over a $\sim 100$ Myrs timescale], which, when taking model and initial conditions for granted,  suggests several changes of apsidal orientation over the age of the Solar system, for this and other TNOs, potentially explaining the opposite apsidal orientation of the likes of 2013 FT28, despite having elements that are consistent with anti-aligned high eccentricity Laplace equilibria. The phase-space shows interplay between Laplace equilibrium islands, and islands harboring Kozai-Lidov like cycling, connected through a sea of chaos signaling resonant interactions induced by the outer perturber. 

\begin{figure}
    \centering
     \begin{subfigure}[b]{0.45\columnwidth}\hfill
         \includegraphics[width=\columnwidth,height=3.7cm]{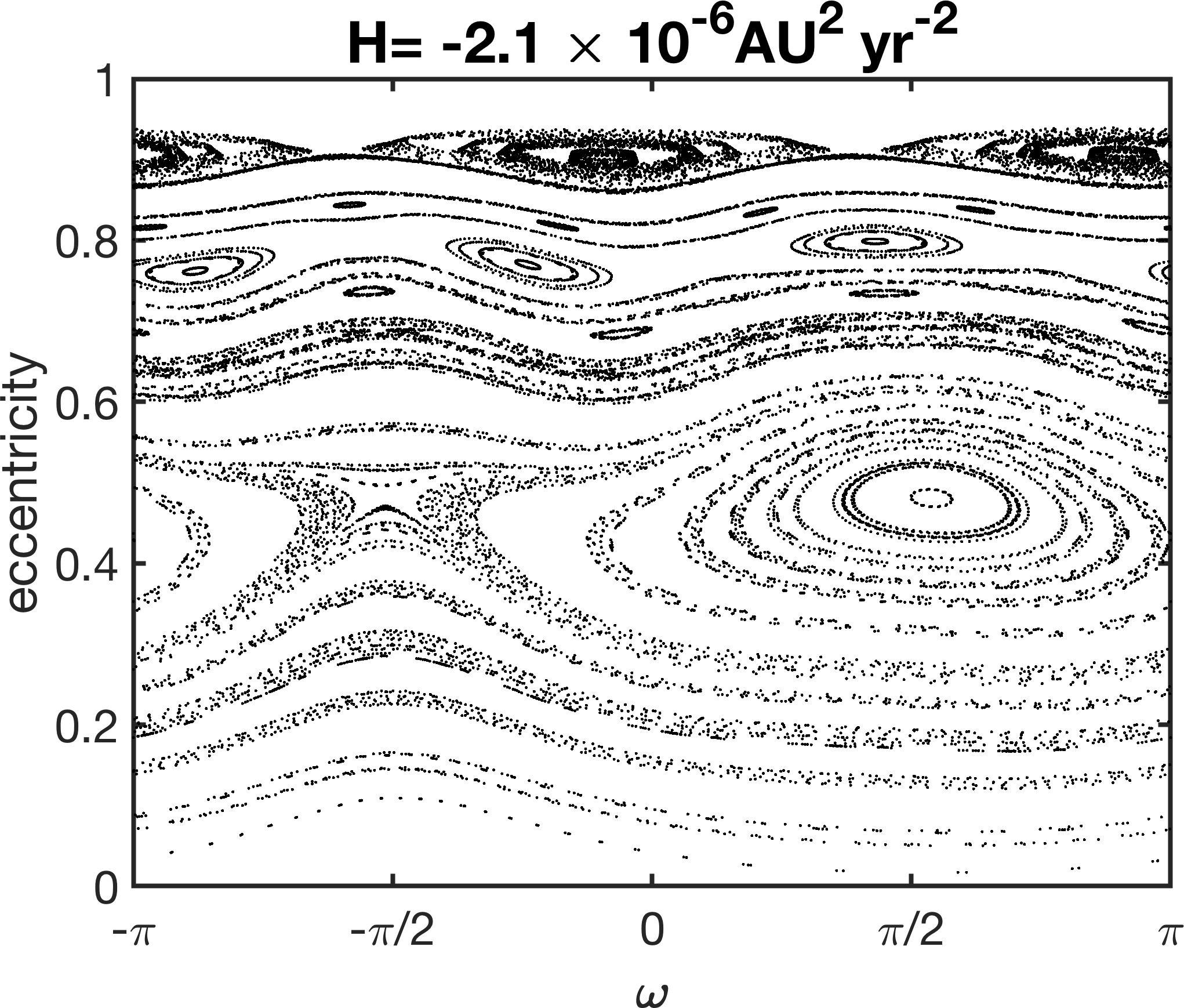}\hfill
    \end{subfigure}  
    \begin{subfigure}[b]{0.45\columnwidth}
         \includegraphics[width=\columnwidth,height=3.7cm]{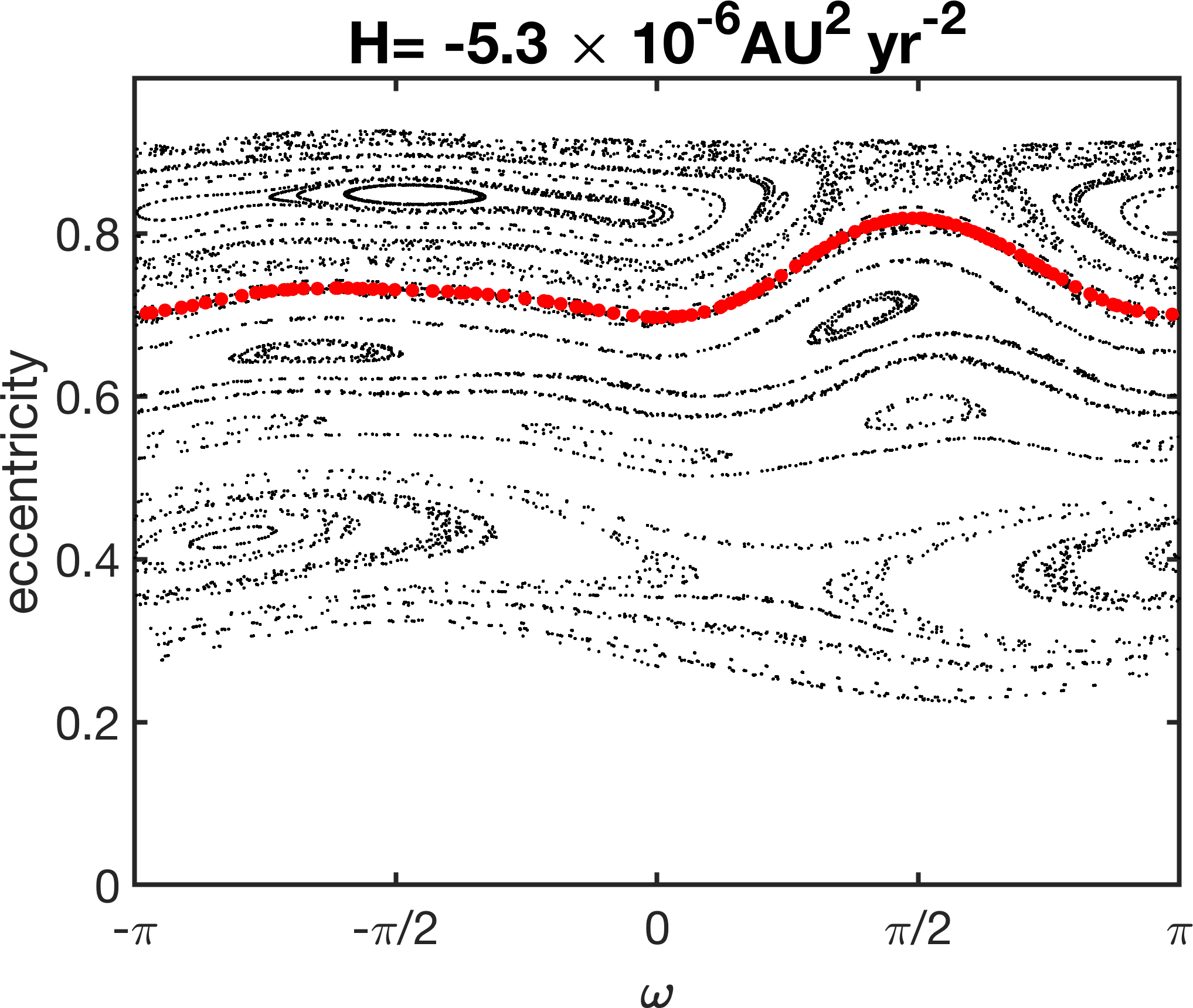}\hfill
    \end{subfigure}  
    \\
    \begin{subfigure}[b]{0.45\columnwidth}
         \includegraphics[width=\columnwidth,height=3.7cm]{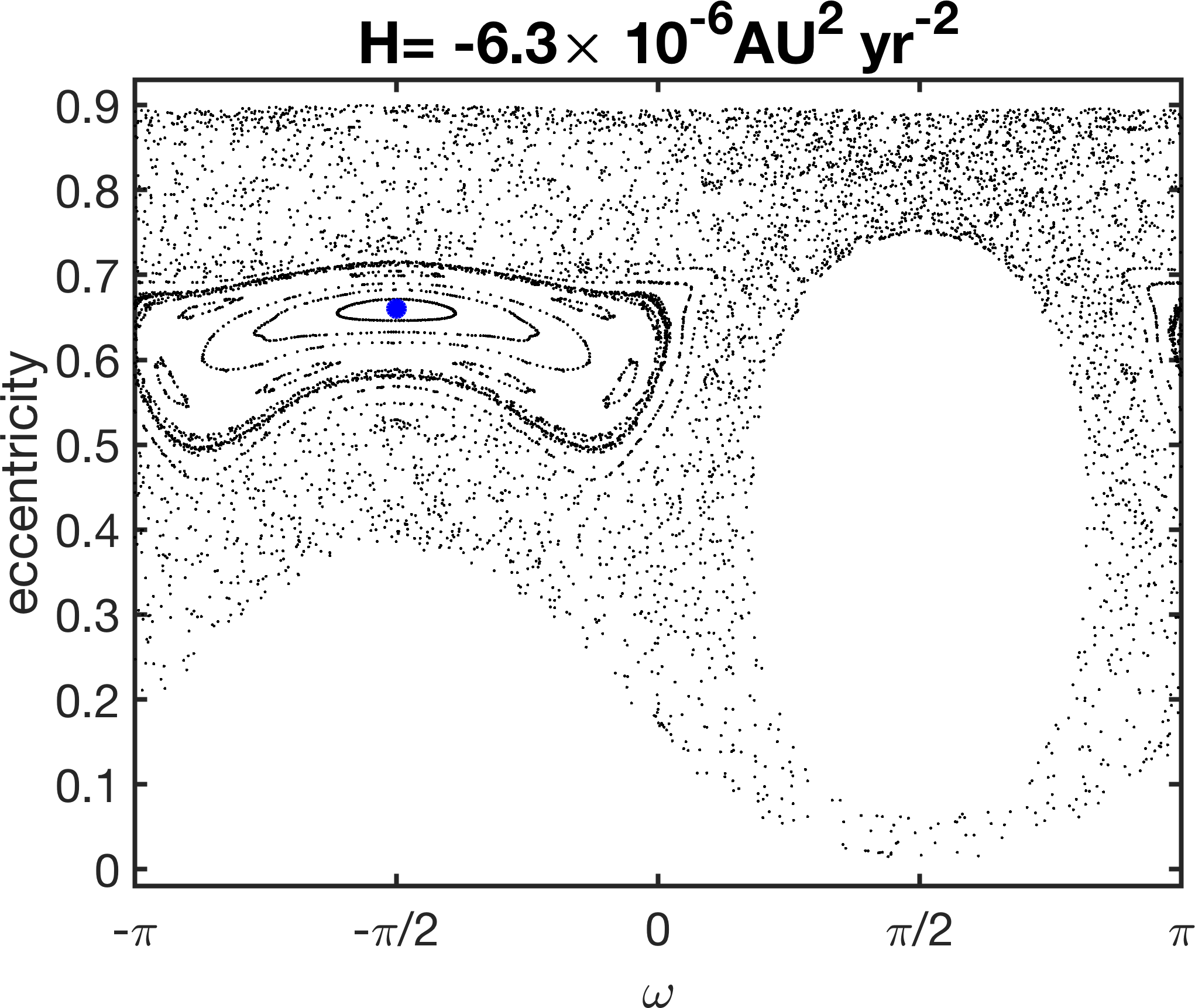}
    \end{subfigure}  
    \begin{subfigure}[b]{0.45\columnwidth}
         \includegraphics[width=\columnwidth,height=3.7cm]{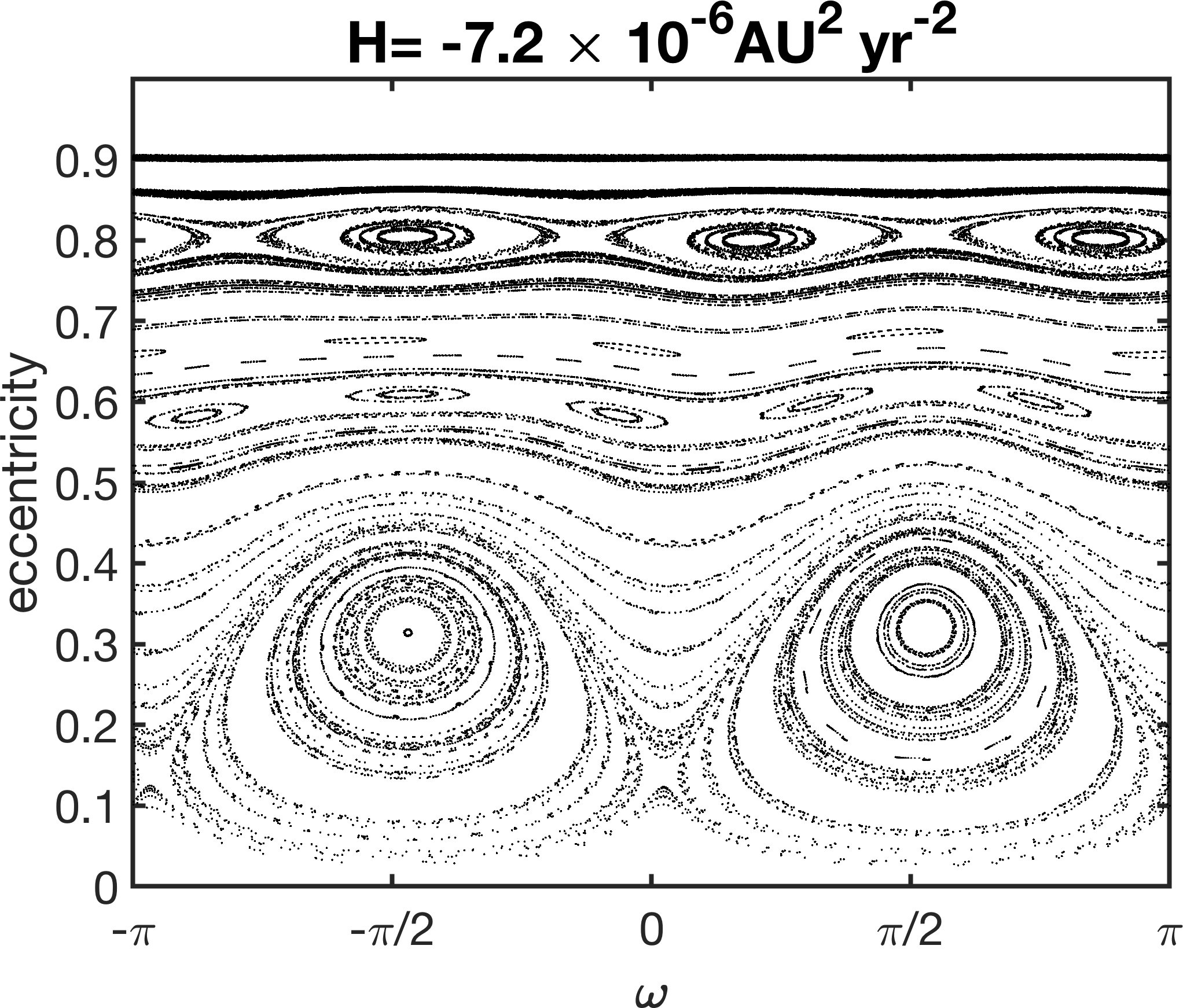}
    \end{subfigure}  
    
    \caption{Same as Fig.\ref{ss-a60}, but at $a=258$ AU corresponding to TNO 2012 VP113. We section at energies straddling the energy of this TNO, passing by the energy of the stable anti-aligned Laplace equilibrium at this semi-major axis. The top-right panel reveals a typical mixed phase space with quasiperiodic motion, tori broken into islands, chaotic zones confined around separatrices...etc. At the lower secular energy of 2012 VP113 of top-right panel, a well defined libration zone emerges around anti-aligned orientations, with the TNO's trajectory (shown in red) hugging it as it circulates between the stable anti-aligned and the unstable aligned fixed points. At a lower energy still, we come across the apsidally anti-aligned Laplace equilibrium at this semi-major axis which is marked with a blue dot in the lower left panel. The libration zone associated with this stable equilibrium is embedded in an extend chaotic zone which connects its neighborhood with that of a Kozai-Lidov island to its right. The lower right panel samples the lowest energy at this semi-major axis, revealing a fully regularized phase space with stable aligned and anti-aligned islands now around $e\approx 0.3$.}
    \label{ss-a258}
\end{figure}

\item $a=302$  AU (Fig.\ref{ss-a302}): The stable high eccentricity anti-aligned equilibrium is tucked within a surviving libration island, which is further embedded in an extended chaotic zone. Shown in the lower left panel is the trajectory of 2014 SR349 describing quasiperiodic motion within the apsidally anti-aligned libration island around a stable periodic center. At the higher energy of $H=-8.9 \times 10^{-7} AU^2 yr^{-2}$, we further code for inclination, revealing how trajectories can transition from aligned, near-circular orbits, at moderate inclination ($~40^\circ$), to anti-aligned, highly eccentric ($~0.8$ ) and fairly inclined orbits ($~80^\circ$). One can then envisage a scenario of decreasing secular energy (at constant $a$), whereby a TNO would transition from the upper to the lower left panel of Fig.\ref{ss-a302}, from a chaotic trajectory at high eccentricity and inclination, to a trapped torus at lower energy and similar orbital architecture. Proto-planetary disk dissipation and/or planetary migration can bring about such a decrease, and with it the trapping of TNOs around the desired eccentric Laplace configuration. 
\begin{figure}
    \centering
     \begin{subfigure}[b]{0.45\columnwidth}\hfill
         \includegraphics[width=\columnwidth,height=3.7cm]{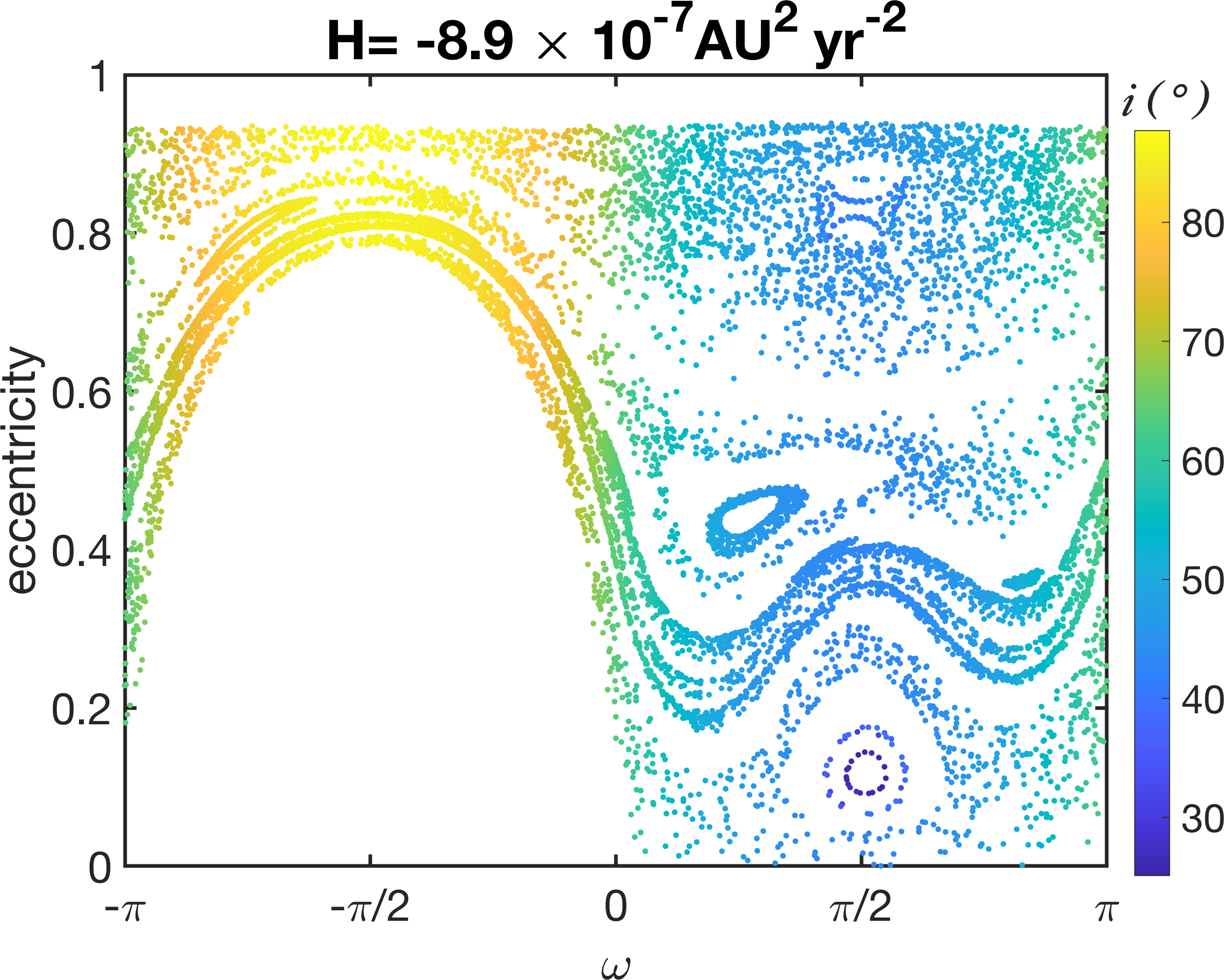}\hfill
    \end{subfigure}  
    \begin{subfigure}[b]{0.45\columnwidth}
         \includegraphics[width=\columnwidth,height=3.7cm]{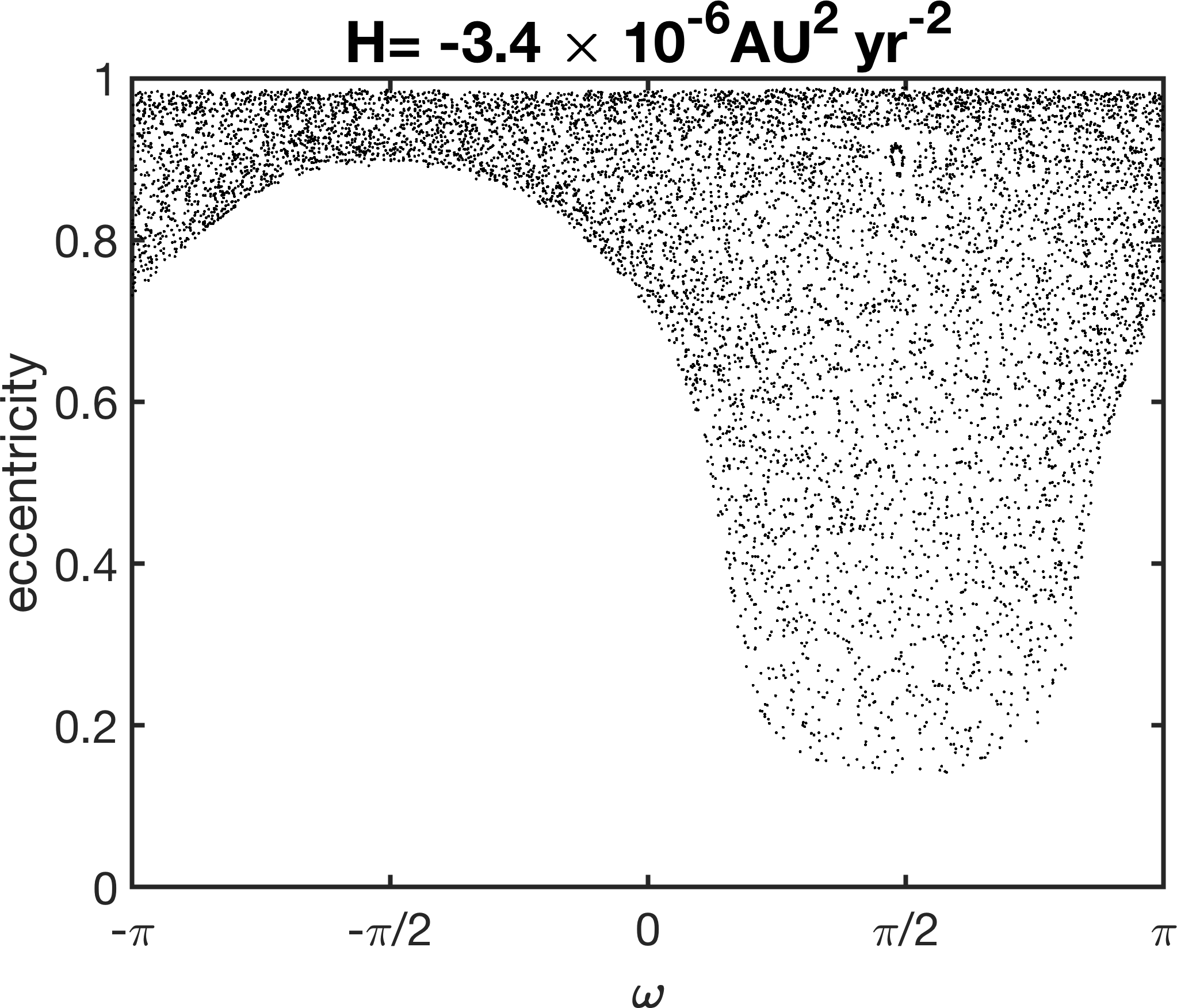}\hfill
    \end{subfigure}  
    \\
    \begin{subfigure}[b]{0.45\columnwidth}
         \includegraphics[width=\columnwidth,height=3.7cm]{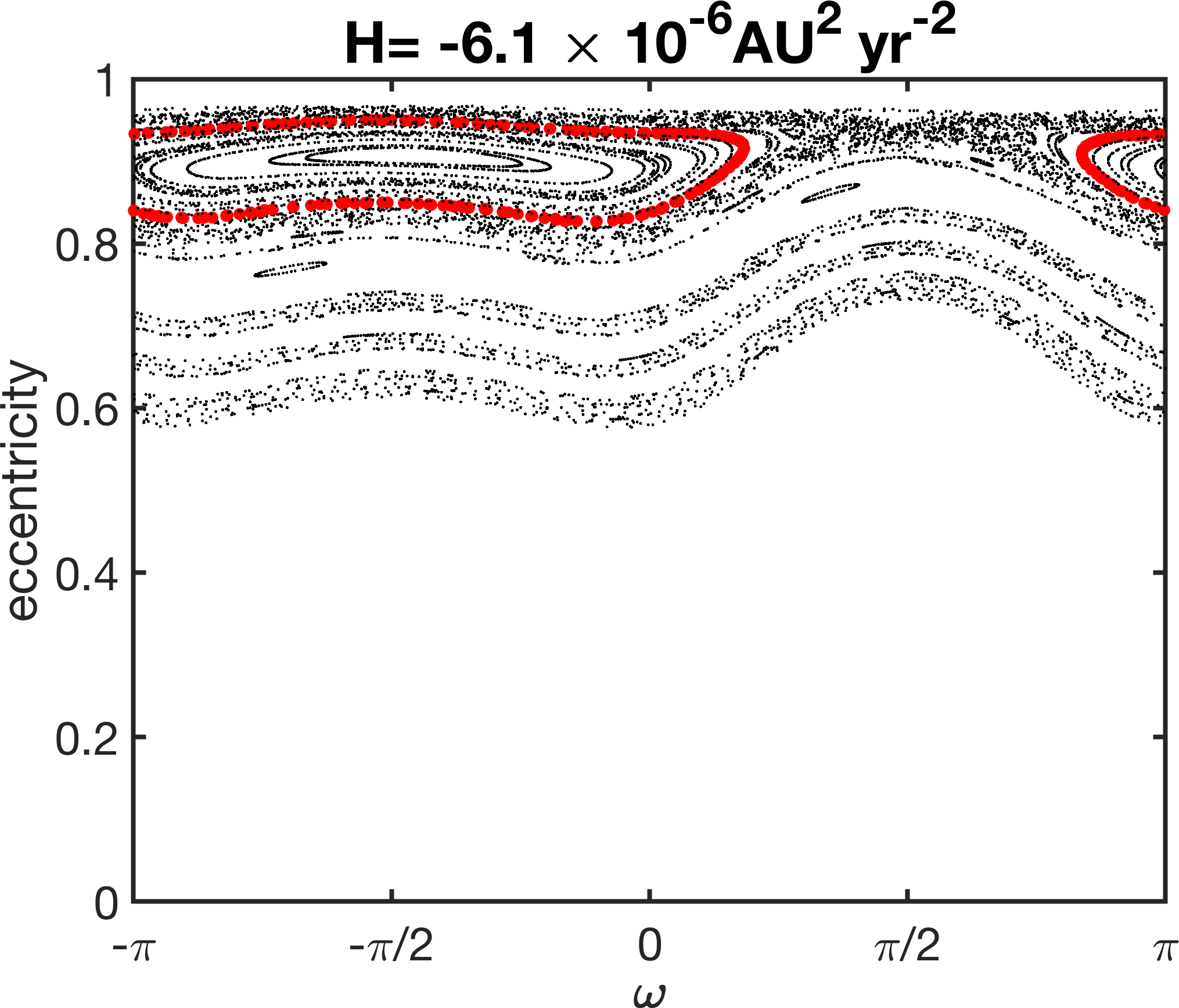}
    \end{subfigure}  
    \begin{subfigure}[b]{0.45\columnwidth}
         \includegraphics[width=\columnwidth,height=3.7cm]{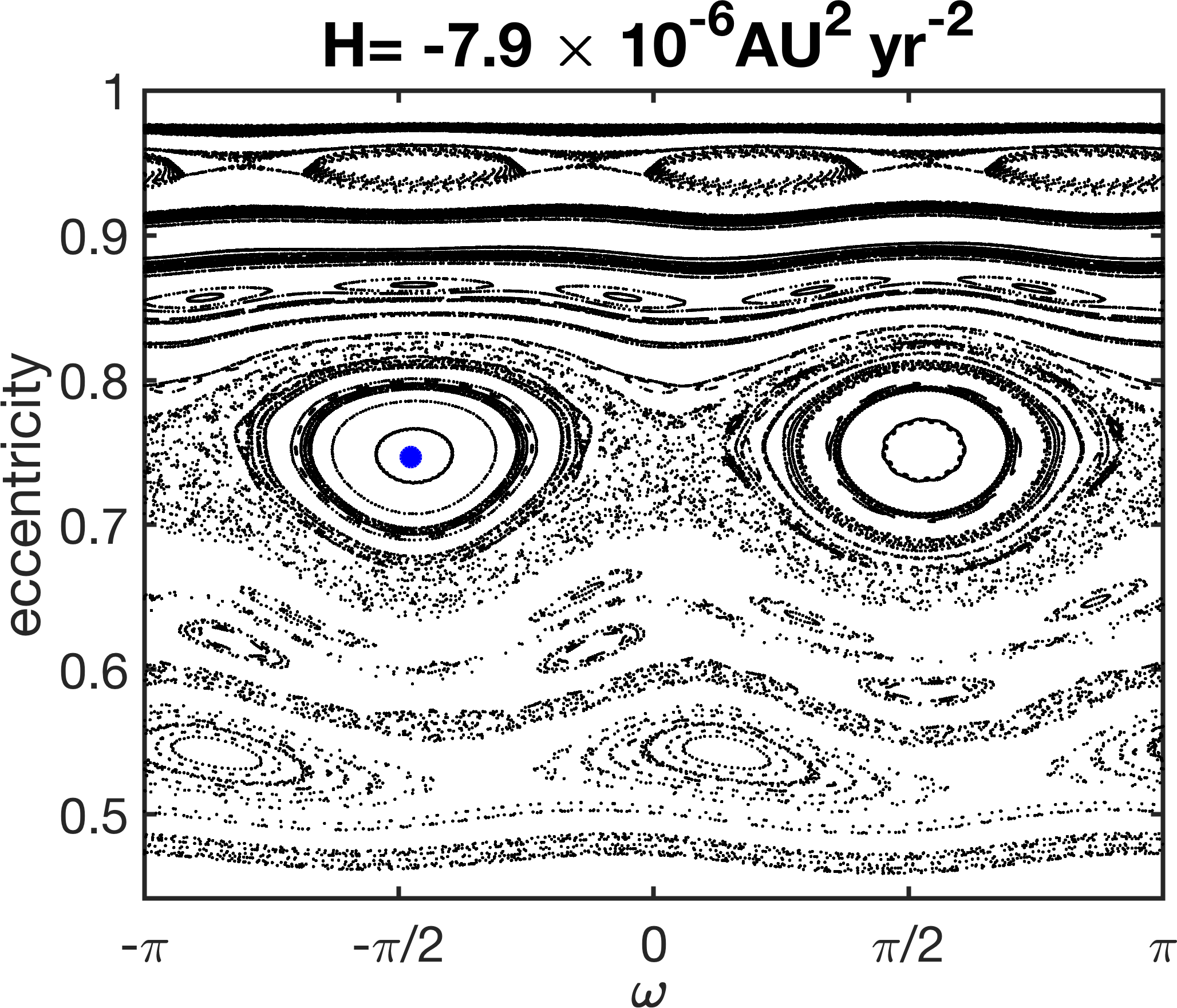}
    \end{subfigure}  
    
    \caption{Same as Fig.\ref{ss-a60} but at $a=302 $AU corresponding to 2014 SR349. The Poincar\'{e} section in the top-left panel is color coded to show inclination variations of the crossings, suggesting possible transitions from a moderate inclination and low eccentricity apsidally aligned zone to high inclination and high eccentricity apsidally anti-aligned zone (see text). Proceeding with the lower energy of the top-right panel, we follow the disappearance of islands at low and high eccentricity, as a broad chaotic zone occupies the allowable phase space. The section in the lower-left panel is computed at the secular energy of the TNO in question. Plotted in red are the crossings of its trajectory, revealing quasiperiodic motion around the anti-aligned island. The section in the lower right panel is computed at the secular energy of the stable anti-aligned Laplace equilibrium, which is shown in blue at the center of the libration zone.}
    \label{ss-a302}
\end{figure}

\begin{figure}
    \centering
    \begin{subfigure}[b]{0.45\columnwidth}
         \includegraphics[width=\columnwidth,height=3.7cm]{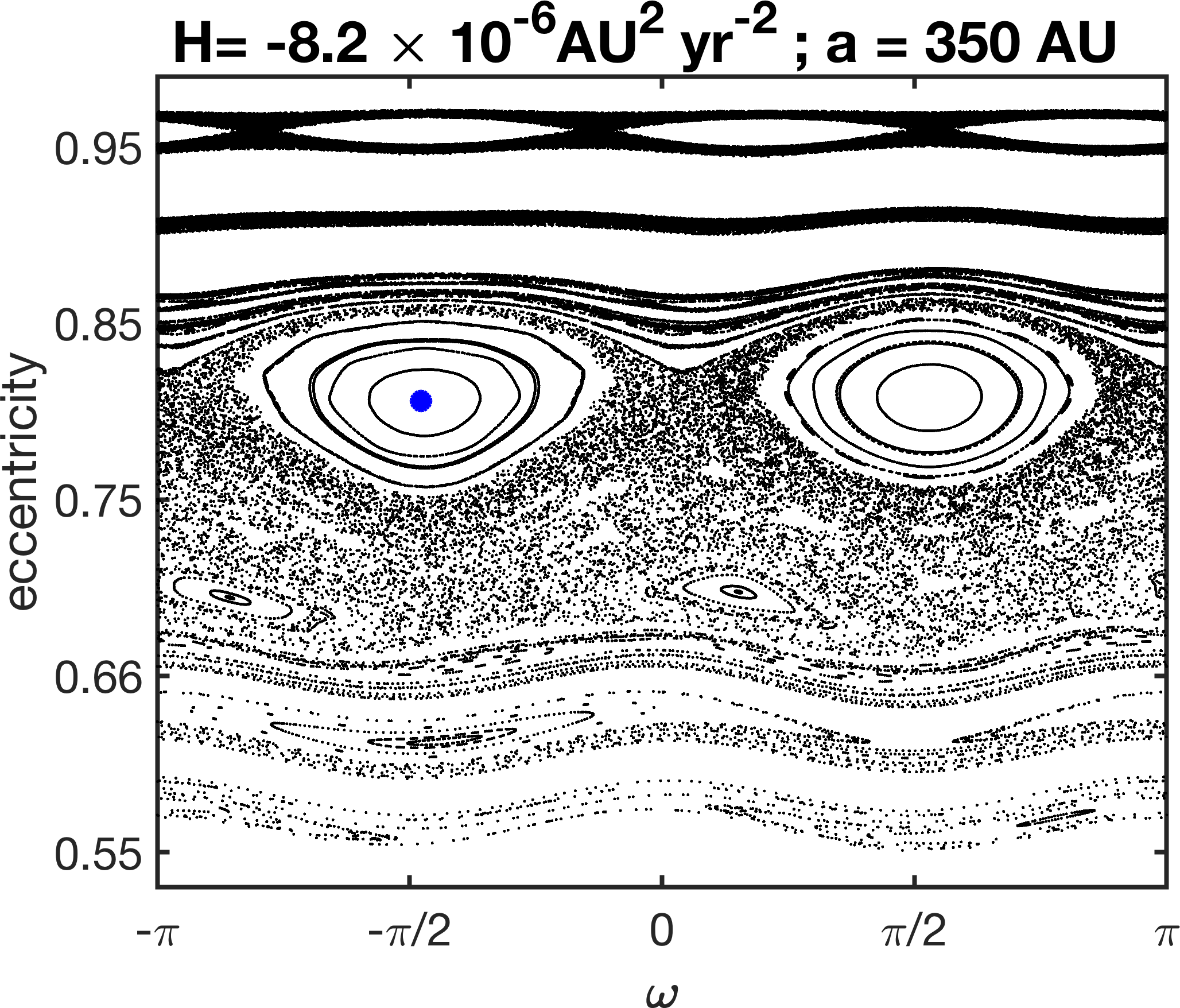}
    \end{subfigure}  
    \begin{subfigure}[b]{0.45\columnwidth}
          \includegraphics[width=\columnwidth,height=3.7cm]{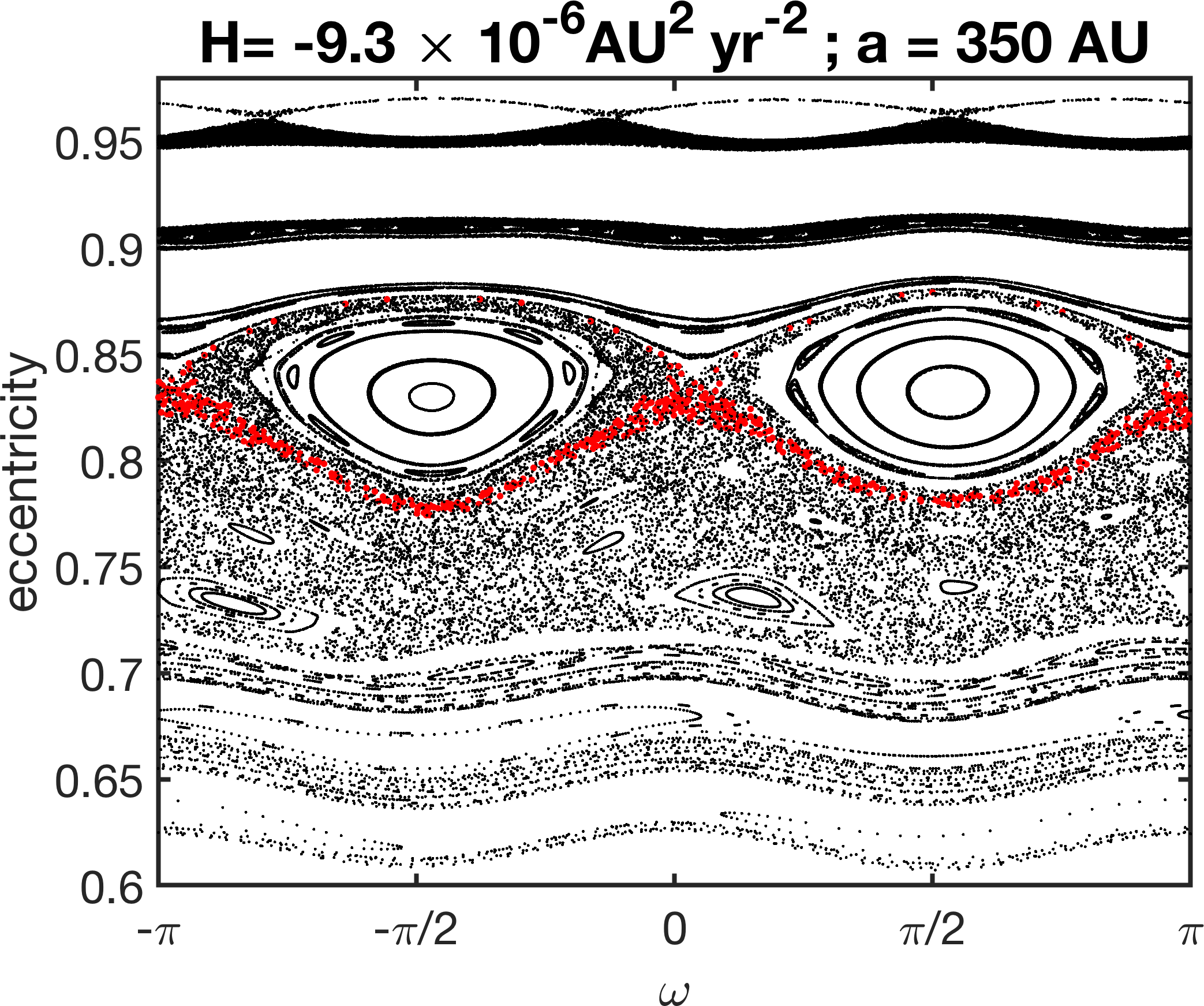}
     \end{subfigure}  
    \\
    \begin{subfigure}[b]{0.45\columnwidth}
         \includegraphics[width=\columnwidth,height=3.7cm]{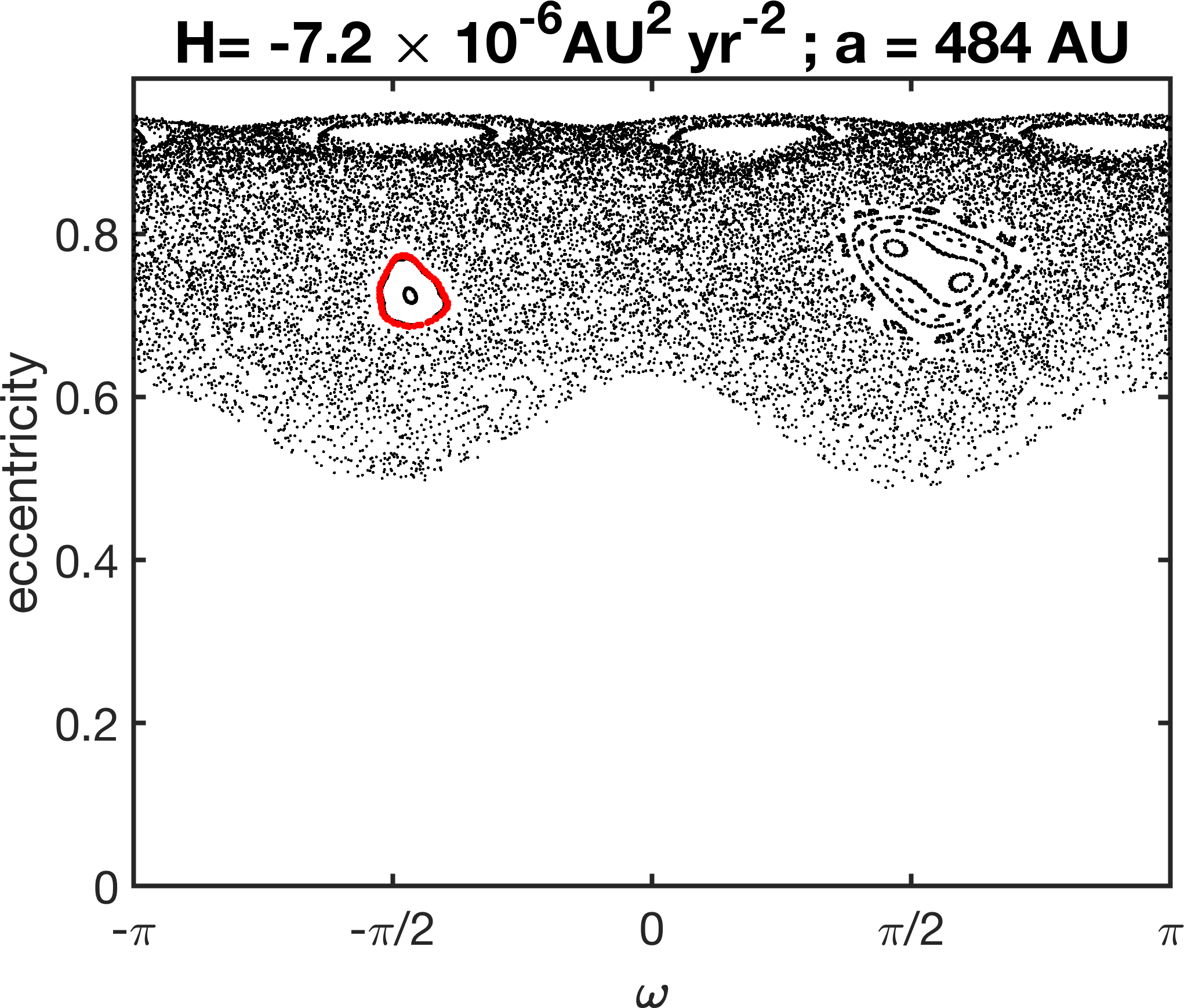}
    \end{subfigure}  
    \begin{subfigure}[b]{0.45\columnwidth}
         \includegraphics[width=\columnwidth,height=3.7cm]{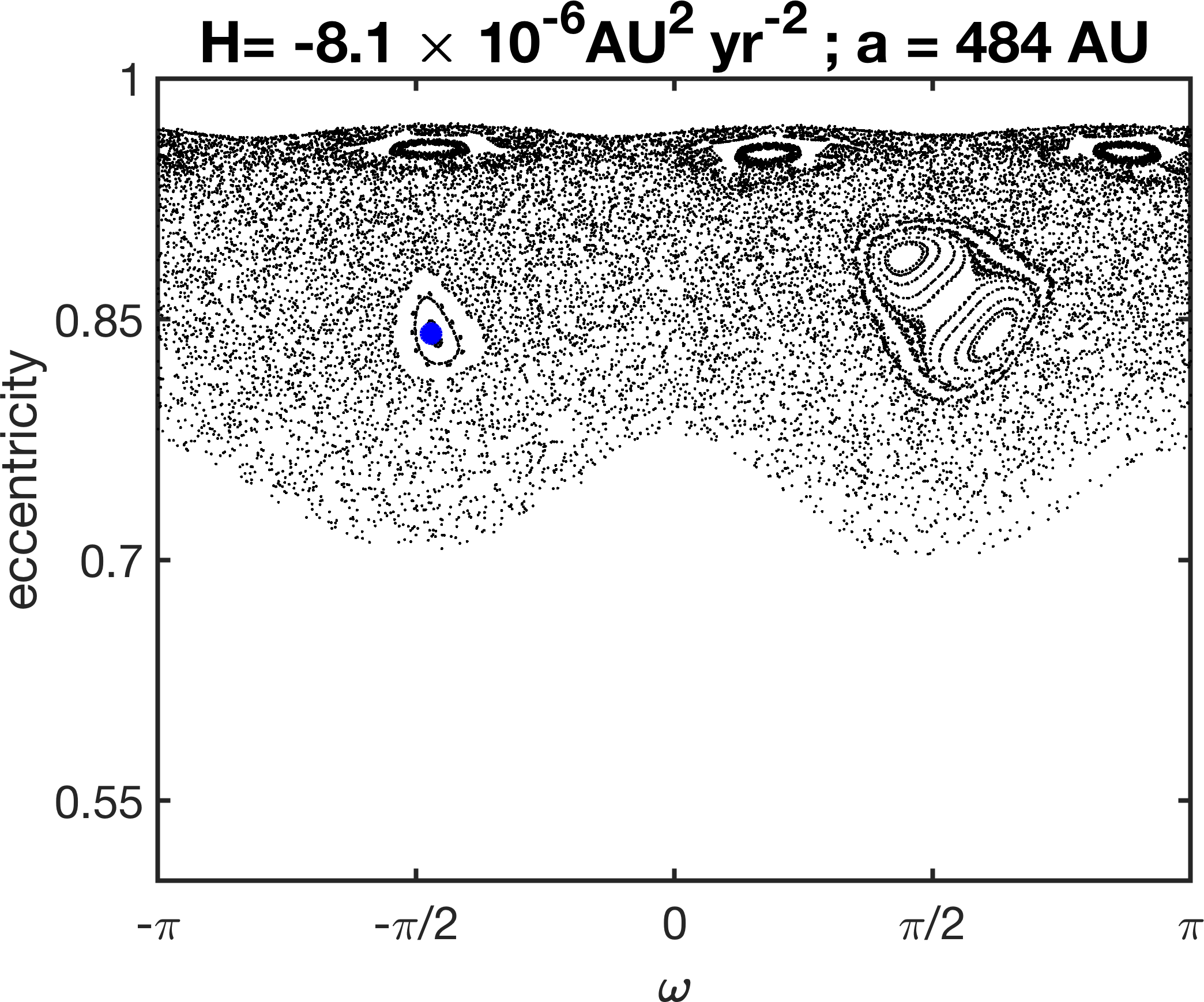}
    \end{subfigure}  
    
    \caption{Same as Fig.\ref{ss-a60} but at $a=350 $AU in the top panels corresponding to 2014 GB174, and at  $a=484 $AU in the bottom panels corresponding to the TNO SEDNA. We section at energies corresponding to the TNOs current trajectories and to stable anti-aligned Laplace equilibria. 
    The trajectory of 2014 GB174 in the top right panel traces chaotic transitions between apsidal alignment and anti-alignment, spending most of its time below the libration islands. Decreasing the energy and moving to the left panel, the phase space structure is largely intact as we locate the Laplace equilibrium centering the anti-aligned libration island. Moving to $a=484 $AU, SEDNA's trajectory inhabits a torus confined within the - now very narrow - anti-aligned libration zone. A slight increase in energy reveals the Laplace equilibrium marked in blue on the right panel. Both sections feature chaotic trajectories occupying the larger area of the restricted phase space.  
     }
    \label{ss-a484}
\end{figure}

\item $a=350$ AU (Fig.\ref{ss-a484}, top panels): We display on the left the stable anti-aligned eccentric Laplace equilibrium embedded within a phase space structure which is similar to that around equilibrium for $a=302$ AU. On the right, the section is computed at the estimated secular energy of 2010 GB174. Its trajectory appears to circulate chaotically between aligned and anti-aligned orientations but still within a relatively narrow chaotic zone, straddling the separatrices, as it hugs the surviving libration zones.

\item $a= 484$ AU (Fig.\ref{ss-a484}, bottom panels): In the left panel, we display the Poincar\'{e} section at the estimated secular energy of SEDNA. SEDNA's trajectory is shown in red, a regular torus which is strictly confined within the anti-aligned libration zone. At this semi-major axis, the main anti-aligned family has reached moderate ecliptic inclinations ($i_E \approx 43^\circ)$. SEDNA's secular energy is actually quite close to the energy of the Laplace equilibrium at that semi-major axis, which is marked on the bottom right panel. Note the narrow allowable range in (high) eccentricity, which is largely occupied by chaos, barring a narrow region of quasiperiodic librations which appear to shelter SEDNA, as well as the stable highly eccentric and moderately inclined Laplace equilibrium. Any scenario addressing the clustering of TNOs must explain their transport and confinement around similar such high eccentricity, and moderately large inclinations.

\end{itemize}

We located eccentric Laplace equilibria within the full phase space available at their, and neighboring, energies. We focused on semi-major axes associated with observed TNOs, and featured their secular dynamics, taking initial conditions as indicators of secular orbital elements within our model. Some appeared caught in finite amplitude stable librations, others to be circulating, on a stable torus in some cases, and around a compact chaotic zone in others. We examined the expected interplay between Laplace and Kozai-Lidov dynamics, as test particles with increasing $a$ move from being dominated by the inner perturber,  to being strongly affected by the outer perturber, with an intermediate zone in between. Though they might appear indistinguishable in the $(\omega,e)$ phase-space, Laplace and Kozai-Lidov libration zones surround equilibria and period orbits respectively, and are connected by transfer orbits transporting particles from near zero eccentricity and small to moderate inclination around Kozai-Lidov cycling, to the high eccentricity and inclination neighborhood of eccentric Laplace equilibria. Such transfer becomes more pronounced with increasing $a$, as aligned and anti-aligned Laplace equilibria increase in inclination, while the inclination of relative Kozai-Lidov equilibria decreases monotonically with a TNOs semi-major axis. Much to explore over this geography, with evident implications for the shaping of TNO orbits by the envisaged P9, but more generally for the sculpting of debris disks by binary companions.

\subsubsection{Case Study: 2015 BP519}
Among the TNOs in Table \ref{eTNOs_table}, Object 2015 BP519 displays curious enough behavior to be featured in this context, as it was in other contexts. Nicknamed \textit{Caju}, it has the highest inclination above the ecliptic ($\sim 54^\circ$), and the closest approach to Neptune ($q \simeq 35.2$ AU). \cite{becker2018discovery} asked two questions concerning the origin and dynamical evolution of this object: $\textit{i)}$ Can it attain it's current configuration starting close to the invariable plane? $\textit{ii)}$ Will it maintain its current configuration in the future? They answer both questions in the affirmative using N-body simulations. We revisit them both within our secular framework. 

\begin{figure}
  \centering
    \includegraphics[width=\columnwidth]{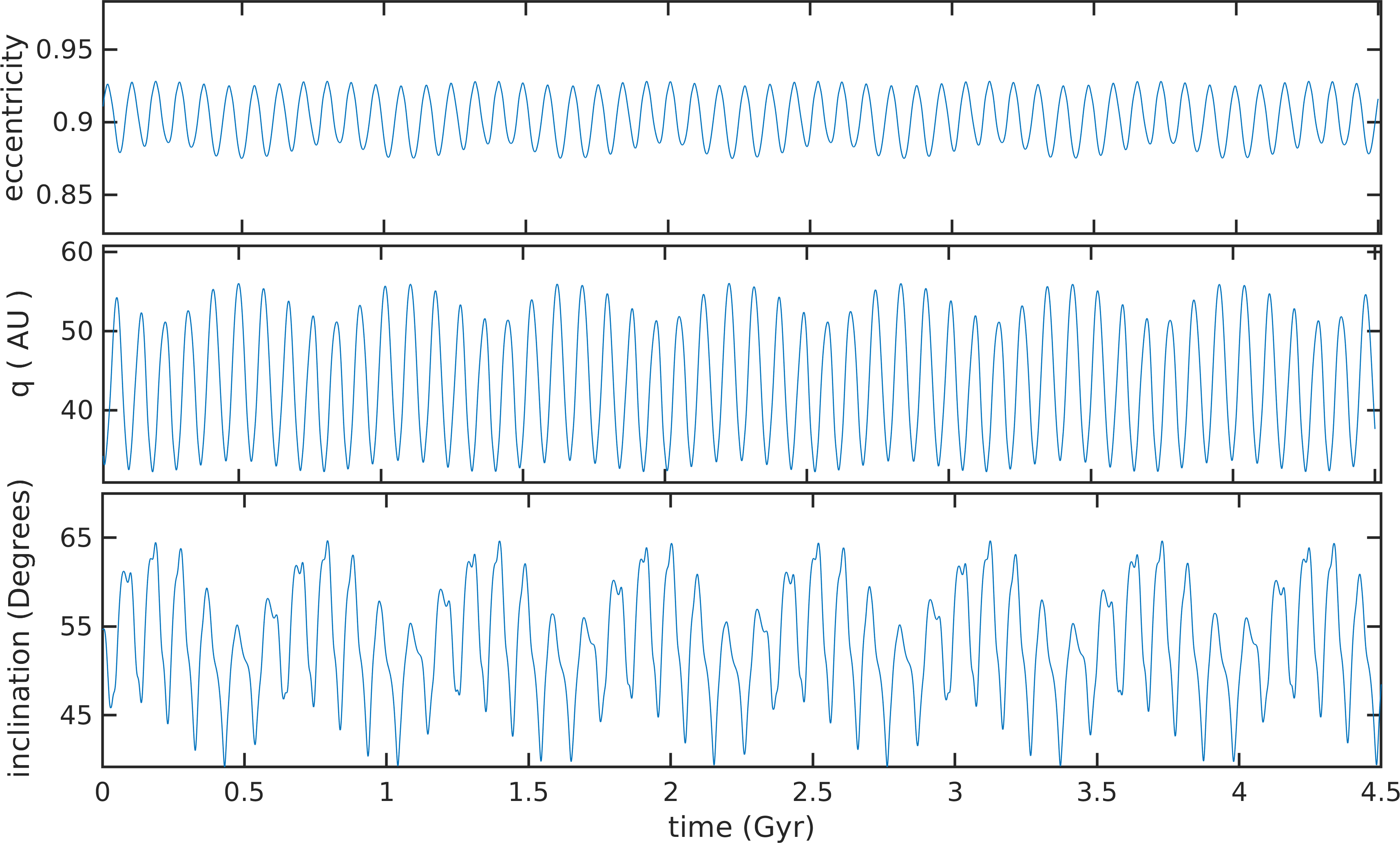}
   
    \includegraphics[width=\columnwidth]{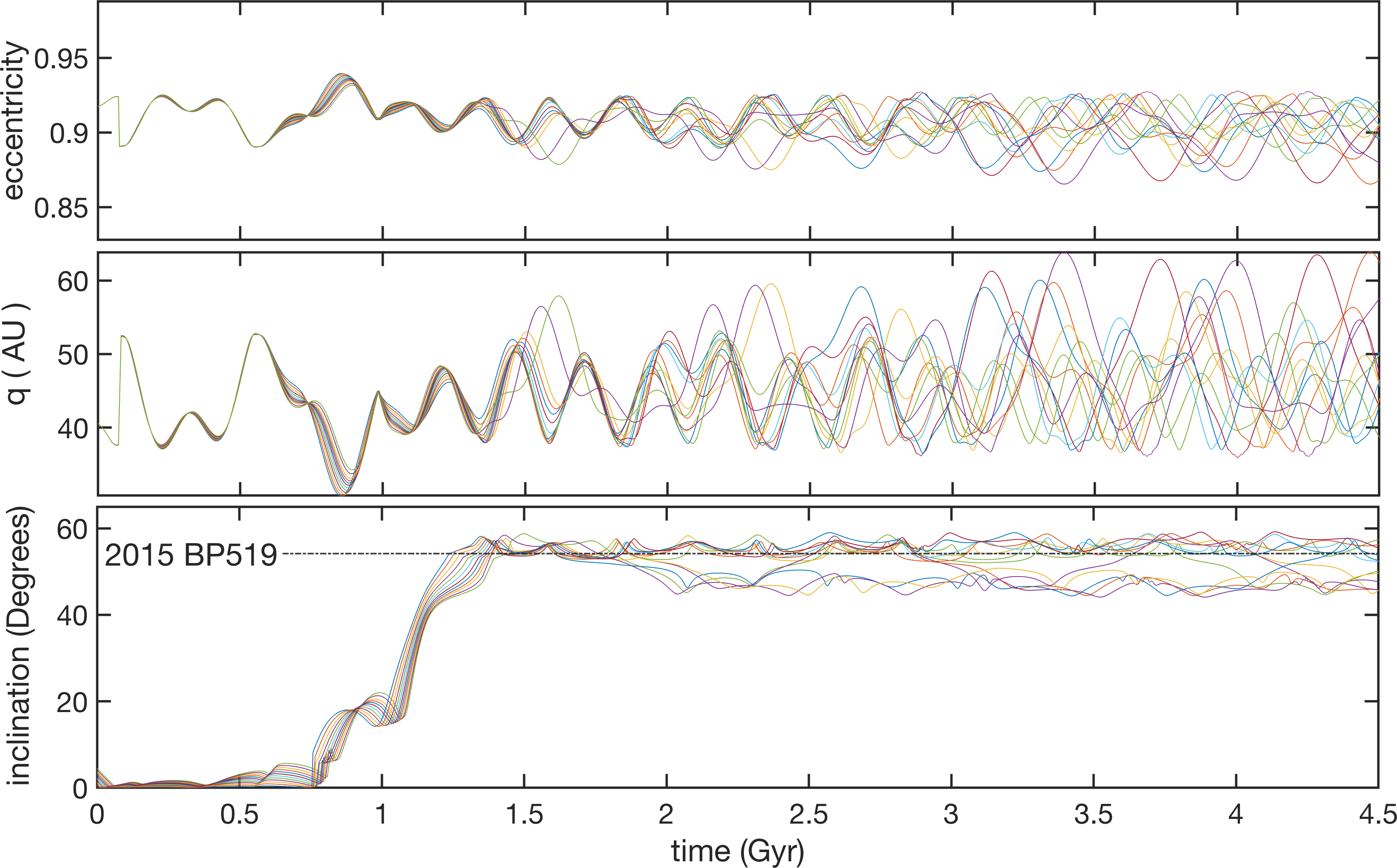}
    \caption{Numerical integration of the equations of motion showing the evolution of the orbital elements of clones of 2015 BP519 under the influence of the giant planets and the putative P9. Top: initial conditions are exactly the elements of the TNO in Table \ref{eTNOs_table}. Bottom: we changed the initial inclination to start around the ecliptic $(0 \leq i_0 \leq 5^\circ$), and the apsidal orientation to start aligned with the outer planet.} 

    \label{BP_evolution}

\end{figure}

\begin{figure}
    \centering
     \begin{subfigure}[b]{0.48\columnwidth}\hfill
         \includegraphics[width=\columnwidth]{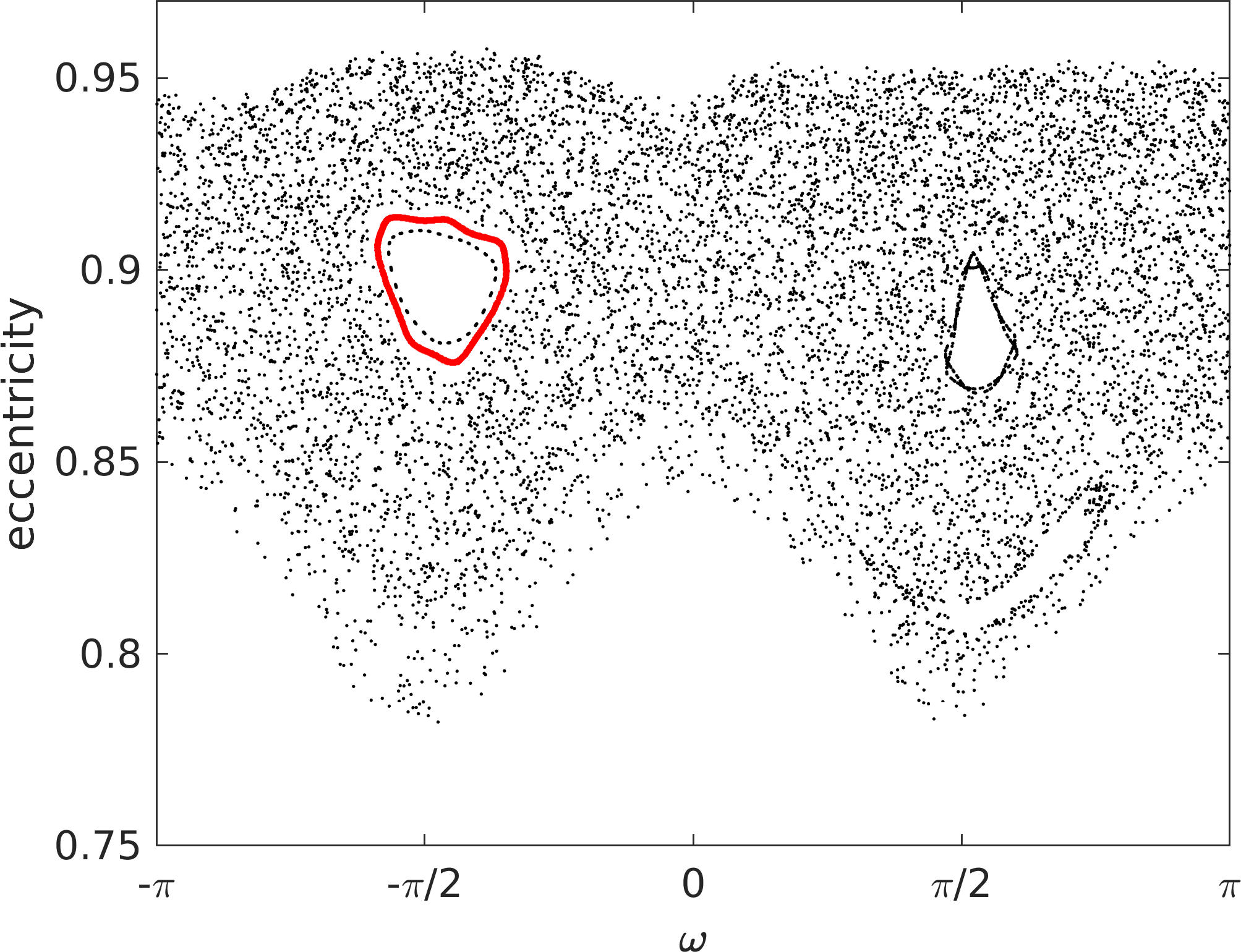}\hfill
    \end{subfigure}  
    \begin{subfigure}[b]{0.48\columnwidth}
         \includegraphics[width=\columnwidth]{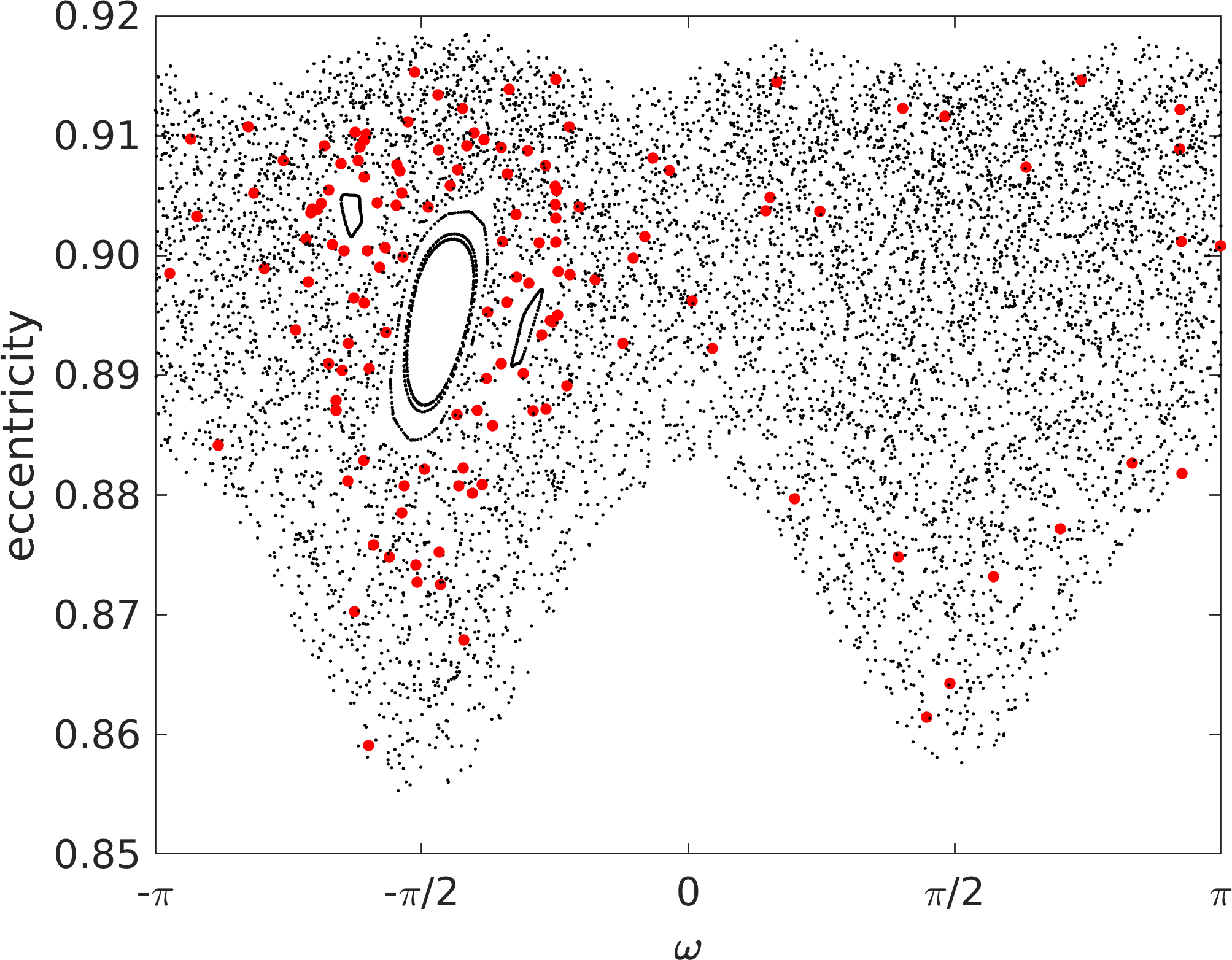}\hfill
    \end{subfigure}

    \caption{Same as Fig.\ref{ss-a60} but at the semi-major axis $a=434$ AU corresponding to 2015 BP519. The first panel is the section computed at the  value of the secular Hamiltonian of this TNO corresponding to its orbital elements as observed and recorded in Table \ref{eTNOs_table}. Its trajectory crossings on the section in red appear to occupy the anti-aligned libration zone. The second panel corresponds to the energy computed for the TNO starting in the ecliptic and around apsidal alignment with P9. }
    \label{ss-BP519}
\end{figure}

In the top panel of Fig.\ref{BP_evolution}, we follow the object over the age of the solar system when initiated around its current orbit. The eccentricity shows small amplitude oscillations ($\Delta e \simeq \pm 0.025$) around $e=0.9$. The inclination behavior is quasiperiodic around its current value, but with relatively large amplitude oscillations, $i=52.5^\circ \pm 12^\circ$. Pericenter distance varies between $q=56$ AU and $q=32.7$ AU. Secular \textit{Caju} appears trapped in and around  its current orbital configuration. In the left panel of Fig.\ref{ss-BP519}, we constructed the corresponding surface of section in $(\omega,e)$ space. At \textit{Caju}'s energy, much of phase-space is inaccessible, with section-crossing trajectories confined to a narrow range in eccentricity, above $e=0.78.$ This region is almost completely filled by chaotic trajectories, except for two evident libration islands around $\omega=\pm \pi/2.$ In red, we trace $\textit{Caju's}$ trajectory, which appears to evolve quasi-periodically within the anti-aligned libration zone. 

We then track the evolution of 12 hypothetical $\textit{Cajus}$, all sharing the same current eccentricity, but with inclinations reset closer to the ecliptic (and picked randomly in the range $0 \leq i_0 \leq 5^\circ)$, and orientation aligned with the outer perturber. As seen in Fig.\ref{BP_evolution}, all particles follow nearly identical low inclination trajectories until about $t=0.86$ Gyrs, when they get unstuck from the ecliptic, with inclinations experiencing a relatively fast period of growth to around the current value. Beyond this point, trajectories start diffusing away from each other and evolving chaotically over a bounded range of fairly high eccentricity and inclination. The surface of section at a typical clone energy (Fig.\ref{ss-BP519}, right panel)  reveals a connected chaotic zone between aligned and anti-aligned orientation, between low inclination and high inclination, and consistently high eccentricity. A sample clone hops around the aligned zone for less than a billion years, then eventually tunnels to the anti-aligned zone, and gets stuck diffusing chaotically around a surviving anti-aligned island. The sharp increase in inclination noted earlier correlates quite neatly with the transition from aligned to anti-aligned chaotic zones. The evolution was extended for twice the displayed duration, with clones remaining trapped around the anti-aligned high inclination zone, with no signs of transition back to aligned, near-ecliptic orientations. 

We discussed how processes which bring about a decrease in secular energy can leave a 2014 SR349 trapped in its orbit, having migrated through its chaotic phase space from an initially circular and moderately inclined orbit. The same process can leave $\textit{Caju}$ trapped in quasiperiodic oscillations, but requires it to start life near the ecliptic, and a fairly large eccentricity. It remains to be seen whether evolutionary processes can, when coupled to rich secular dynamical evolution, bring about anti-aligned or aligned clustering at high eccentricity and inclination, with TNOs starting around the ecliptic with relatively small eccentricity. 

%% file: Discussion.tex
\section{Discussion}
Before concluding, some remarks putting our work in context, drawing implications, reporting on extensions: 
\begin{itemize}
    
     \item \textit{On Inclined Equilibria:} We explored eccentric inclined equilibria with TNOs in mind, but our results may very well be relevant to exoplanetary systems. In that context, variations in the inner quadrupole \citep{spalding2016spin, spalding2018resilience}, then  giant planets residing in wide inclined orbits \citep{pu2018eccentricities} have been shown to excite mutual inclination in multiplanet systems. We combined both effects and showed how they annihilate on eccentric and inclined orbits on the \textit{Eccentric Laplace Surface}. The fictitious system of Sec.\ref{ecc.lapl.surf} is not unlike Kepler-108 \citep{rowe2014validation}, with an inner planet, Kepler-108 b,  serving as the inner quadrupole, and the stellar companion, Kepler-108 A,  providing the inclined and eccentric wide perturber, both torquing the planet between them, Kepler-108 c. The observed mutual inclination of $24^{ +11 \circ}_{-8}$ \citep{mills2017kepler} is consistent with results  reported in Figs.\ref{E-LS} and \ref{multipoles_IB50}, though further investigation is required due to the proximity and comparable masses of the planets.
 
      \item \textit{On P9 alternatives:} \cite{sefilian2019shepherding} argued that a moderately eccentric $(e\approx0.2)$ disk extending between 40 and 750 AU with a mass of $10M_\oplus$ can together with the giant planets shepherd TNOs into coplanar apse aligned configurations . They further suggested that the combined action of such a disk, the giant planets, and a scattered inclined planetary embryo [outcome of simulation in \citep{silsbee2018producing}] might be enough to maintain TNOs into inclined apse aligned configurations. This of course smells like the eccentric, inclined spatially frozen Laplace equilibria of our theory. Our preliminary explorations of this setting involved generalizing the modal structure from one ring describing P9 to N-rings describing a self-gravitating precessing debris disk. Combining the action of this disk with the giant planets and an inclined planetary embryo as small as Mars, we could recover families of eccentric and moderately inclined equilibria that match the orbital trends of the observed TNOs without the need for a massive outer planet. 

   \item \textit{P9 and TNO Clustering:} In this work, our intention is not to engage critically with the P9 hypothesis. Rather, we simply argue that eccentric inclined Laplace equilibria in the combined field of the gaseous giants and a putative P9 are those desired frozen orbits around which the TNOs can be clustered in librating apse-aligned configurations. We argue along those lines, and proceed to solve for this skeletal structure, then map the chaotic secular phase space within which it is embedded.  
    
    We showed how the planar structure of equilibria is smoothly transported into an off-plane structure upon tilting the orbit of P9. Libration islands around equilibria could harbor apsidal anti-aligned clustering, as shown specifically for 2014 SR349, 2015 BP519, and SEDNA. The apsidal orientation of the likes of 2013 FT28 and 2015 KG163 can be further explained by the emergence of a highly eccentric aligned family or trajectories surfing both apsidal areas of the Poincar\'{e} sections.
    
    Furthermore, one notes that clustering obtains through confinement of trajectories  to compact chaotic zones in phase space, for a broad range in energy [similar behavior was reported in \cite{saillenfest2017non}].  Such confinement is often accompanied with circulation between anti-aligned and aligned configurations, typically over a $~100$ Myr timescale. Thus,  it should not be surprising to find objects with the high eccentricity an inclination of an anti-aligned Laplace equilibrium, spatially aligned with the outer-perturber's periapse. The more pressing question for any similar such scenario for TNOs clustering  is how to get them trapped out there in the first place! 
    
    We leave it to P9 enthusiasts to consider and assess evolutionary scenarios over the (secular)-dynamical landscape. This said we have highlighted promising cases where chaotic exchange orbits between Kozai-Lidov and Laplace-like regimes, can provide a vehicle for transporting TNOs close to high eccentricity/high inclination Laplace equilibria. A slight tip in energy can leave that same object trapped around its neighboring Laplace equilibrium.

    Our secular formalism does not allow for semi-major axis diffusion, which seems tolerable given how the semi-major axis time-series of most of the objects presented in \cite{batygin2019planet} showed no marked diffusion. It further excludes confinement due to mean motion resonances with the gaseous planets, or disruption of secular equilibria by those same resonances: it would be interesting to explore perturbations of idealized secular equilibria by terms in the disturbing function which are of second order in the masses. 

\item \textit{Equilibria and Slow Nodal Regression:} Fixing a perturber's orientation reveals dynamics for a specific configuration, at a particular point in time, whereas we know that both outer and inner perturbers precess, slowly, but they precess nonetheless. Indeed a P9-like object is expected to precess with a 10 billion year time scale, and force the inner quadrupole into precession over a 100 billion year time scale. Contrast this with the expected secular dynamical timescales of TNOs [two to three orders of magnitude faster, with typically a 100 Myr precession period], and you can see why we are fully justified in working in the adiabatic limit \citep{arnol2013mathematical}. In that limit, equilibria are expected to remain close to equilibria, and quasiperiodic motion to deform into quasiperiodic motion, with an adiabatically invariant action, suitably defined over the surface of section. Those expectations were confirmed in extensive simulations exploring dynamics in a sequence of Hamiltonians, corresponding to peturbers which are frozen in a sequence of relative orientations. Libration zones around stable Laplace equilibria were seen to oscillate as they deform adiabatically with the circulating quadrupole. This was so over a range of semi-major axes, suggesting that equilibrium families, rather than being tuned to the particular orientation of Fig.\ref{3d_Equlibria}, are largely preserved as they deform adiabatically, reversibly, with quasiperiodic adiabatic changes in the perturbers. Preliminary results to be sure, and deserving of further exploration, but encouraging nonetheless.

\item \textit{On Weakly Hierarchical Systems:} Our method for dealing with such systems has much promise, and deserves to be further confronted with costlier though more accurate numerical double averaging  \citep{beust2016orbital, batygin2017dynamical}. The astute reader may wonder about the structural stability of the equilibria we recovered with this approach, when one increases the order of harmonics included in the potential of the outer perturber. We wondered about the same and learned, that having passed the convergence test, harmonics of order higher than the third considered in our work, produced an almost identical skeleton of equilibria, confirming the robustness of our results and their reliability as foundation for further studies. 

We conclude this segment by noting that the method we presented is ideally suited for fixed, or uniformly precessing configurations, and can be generalized to any number of particles (modeled as Gaussian rings), as was recently demonstrated in  \cite{sefilian2019shepherding}. Of course, the method of choice to fully capture secular test particle evolution in dynamically evolving (inner and/or outer) perturbers, is the softened Gauss algorithm \citep{touma2009gauss} which we hope to deploy on eccentric Laplace dynamics in the near future.

\end{itemize}

%% file: Conclusion.tex
\section{Conclusions}
We examined the orbital architecture of test particles under the combined effect of an inner and an outer perturber residing on an inclined and eccentric orbit. Classically, inner and outer perturbers are included up to quadrupolar level, and the problem reduces to finding stationary test particle orbits on which the secular perturbation of the combined quadrupoles vanishes \citep{Tremaine}. Critical to this story is the so called \textit{Laplace Surface}. Traced by a distinguished family of circular Laplace equilibria, this warped surface coincides with the plane of the inner perturber close in, and  transitions to the plane of the outer perturber at large distances. It has played a fundamental role in the understanding of planetary satellites, our Moon included, and has experienced a dramatic revival as theorists grapple with an influx of complex hierarchical architectures, from the plethora of exo-planetary systems on one scale, to stellar black hole nuclei on another. 

Our work generalizes the already productive generalization of \cite{Tremaine} by including higher order, symmetry breaking perturbations from an eccentric outer perturber, all the while following the authors' lead in exploring equilibria in arbitrary architectures. Our generalization is meant to bring Laplace Surface dynamics closer in touch with: exoplanetary architectures with highly eccentric wide binary companions; satellite dynamics around planets undergoing eccentric Kozai-Lidov cycling; secular dynamics of accretion disks and stellar clusters with an eccentric binary black hole; and/or debris disk architecture with gaseous giants tucked in, and a massive eccentric perturber, be it an outer planet, or the disk itself, or both.

Focusing on circular equilibria, it is perhaps not surprising to learn that the inclusion of an outer octupole shatters the classical \textit{Laplace Surface}, leaving an \textit{Eccentric Laplace Surface} on its ruins. This surface is parametrized with eccentric, inclined, equilibrium orbits, evolving in orientation in a manner analogous to their classical counterparts, with eccentricity increasing as they approach the outer perturber. Interestingly enough, this surface is accompanied by the bifurcation of stable, retrograde, and highly eccentric families of equilibria. 

We highlight the significance of those novel equilibria for various current astrophysical settings. In an exo-planetary system, where the inner quadrupole is provided by a compact coplanar system, and the outer octupole by an inclined and eccentric outer massive planet or a stellar companion, the \textit{Eccentric Laplace Surface} may help explain observed  mutual inclinations and eccentricities of architecturally complex systems like Kepler-108 and Kepler 419 \citep{rowe2014validation, dawson2014large}. 

Closer to home, it was natural to examine the frozen orbits of our generalized Laplace equilibria as natural parking (phase)-space for TNOs under the combined gravitational perturbations of the outer planets and a hypothetical 9th planet \citep{trujillo2014sedna, batygin2019planet}. The setting being weakly hierarchical, trusted multipole expansions are hopelessly divergent over a wide range of relevant semi-major axes. Previous studies dealt with this problem through costly "exact" numerical averaging of the Hamiltonian \citep{beust2016orbital, saillenfest2017non}. We overcame this hurdle with a reasonably efficient and accurate fix, capturing the averaged potential of an outer perturber (thought of as a Gaussian ring of arbitrary eccentricity and orientation à la \cite{touma2009gauss}), through spherical harmonics which are then averaged over an arbitrary test particle orbit. 

Suspending concerns about observational bias \citep{shankman2017ossos, brown2017observational, napier2021no}, we use  our tested and versatile toolbox to show how families of eccentric, inclined and stable Laplace equilibria maintained by a fiducial 9th planet are strongly correlated with the phase-space distribution of the TNOs which that planet is expected to shepherd. We thus confirmed our hunch on eccentric, high-inclination Laplace clustering, further providing modelers with a skeletal structure of secular equilibria around which to elaborate variations with non-secular effects, should they so desire. 

Equilibria were further situated within the broader phase space as we mapped global dynamics with suitably constructed  Poincar\'{e} sections, at TNO semi-major axes, and with energies close to our best estimate of TNO secular energy. Stable equilibria emerged at the centers of libration zones, themselves potential trapping zones for clustered TNOs. Chaotic dynamics was shown to engulf the available phase space volume with increasing TNO semi-major axis. We further gave evidence of stickiness within chaos, providing long term confinement of TNOs, diffusive though it may be. With global structure in hand, we explored the evolution of the highly inclined TNO 2015 BP519. Using current conditions, we reveal its confinement  as it librates around the neighboring Laplace equilibrium; then allowing it to start life closer to the ecliptic, we showed how (in agreement with \cite{becker2018discovery}) it can evolve then stick and diffuse around its current inclination. We further revealed the coming together of two secular dynamical features, Laplace and Kozai-Lidov, with Laplace equilibria growing in inclination as we approach the outer perturber, and Kozai-Lidov zone decaying to settle into low inclination regimes. This sequence of events, the geometry that underlies it, is further shown to be robust to slow nodal precession of the inner quadrupole, an adiabatic regime in which equilibria and associated islands survive as they undergo slow periodic shifts with the periodically varying orientation of perturbers (inner and outer alike).

In discussion, we highlight various implications and variations. In particular, and building on a proposition of \cite{sefilian2019shepherding}, we reported preliminary results on how an extended and moderately eccentric precessing debris disk can join forces with an inclined planetary embryo, as puny as Mars, to self-consistently generate a structure of equilibria that also matches the trend of orbital parameter of the TNOs, without the need for an extra massive planet. 

So where does all this leave us? Well, the last major generalization of Laplace's work ushered a stream of applications, mainly to extra solar settings \citep{munoz2015survival, zanazzi2016extended}, but also to scenarios of Lunar formation which argue for an initially steeply oblique and fast spinning Earth \citep{cuk2016tidal,tian2020vertical,cuk2021tidal}. We can foresee the same for our renewed focus on the foundations of the Laplace Surface itself, and for the rich structure of equilibria we have identified, with applications ranging from man made satellites, to exo-moons around planets on Kozai-Lidov cycles, then debris disks with enclosed planets, the whole perturbed by a wide eccentric binary. A key missing ingredient in our story, one which was alluded to in discussion, concerns evolutionary processes (proto-planetary disk dissipation, instabilities in multiplanet systems, planetary migration, disk relaxation and potential instabilities encountered on the way) which can then allow for dramatic events over the phase space structure of our work: loss of stability through bifurcations, capture and evolution along a specific family of equilibria, transition between Kozai-Lidov and Laplace regimes, and the implication of those phenomena for the carving of distributions of particles, including the emergence of clustering, should that be of interest! Within the octupolar limit, we have barely managed to map out one class of equilibria, the so called coplanar-coplanar variety, and drew out its consequences for various astrophysical applications. Considered in full generality, our dynamical systems sustains additional, orthogonal, equilibrium geometries which are worth mapping in full detail, and again with various potential applications in mind (refer to \cite{Tremaine} for further discussion). Finally, having developed machinery for the harmonics of a single ring, it would be natural to deploy it  on a distribution of such rings, perhaps one representing the lopsided thermodynamic equilibria of self-gravitating secular disks, in the presence of an inner quadrupolar perturber, be it a a supermassive black hole binary, or a planet or a system of such planets.

%% file: A1.tex
\section{Orbit Averaged Potential of the P9 Gaussian Ring}
\label{App_P9_potential}
\begin{figure*}
	\centering
	  \includegraphics[width=\textwidth,height=3.2in]{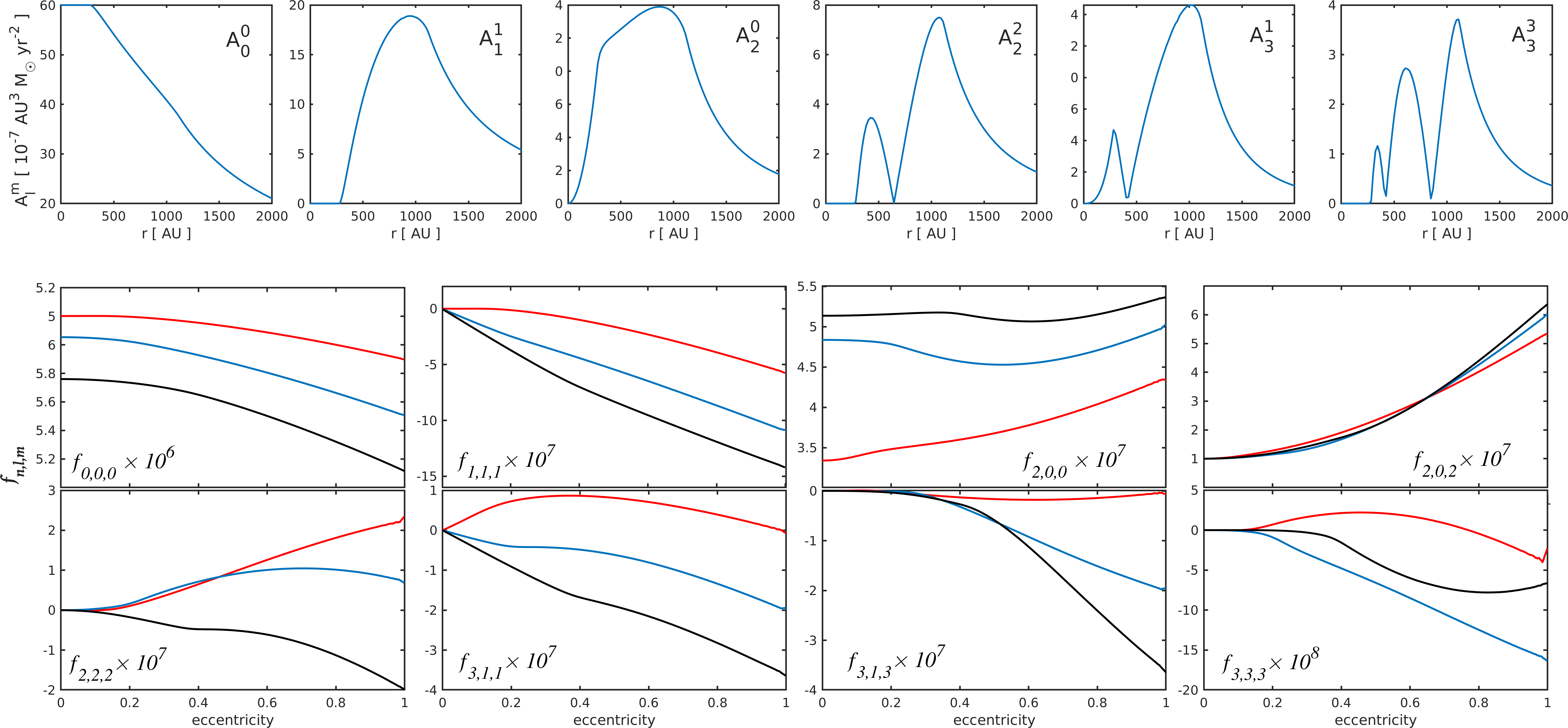}
	\caption{Orbit Averaged Harmonics and the potential of the P9 ring. Top row: Absolute values of the modes $A_l^m(r)$ used in the expansion of Eq.\eqref{p9_potential_modes}. The modes decrease in amplitude as their harmonic degree increases, from the mostly dominant $A_0^0$ mode associated with the axisymmetric contribution, to higher harmonics corresponding to spatial asymmetries. Bottom rows: A sample of the numerically averaged functions $f_{l,m,n}(a,e) = \langle A_l^m(r) \cos(nf)\rangle$ used in the averaged potential in Eq.\eqref{p9_potential_harmonics}. Each function is plotted for three values of the semi-major axis, namely: $a=$ 250 AU (Red), 350 AU (Blue), 450 AU (Black). }
	    \label{modes_fig}
\end{figure*}

The gravitational potential of the Gaussian ring associated with P9 is computed numerically over a three dimensional grid, then expanded in spherical harmonics. Allowing for the dominant non-axisymmetric modes (shown in Fig.\ref{modes_fig}), together with the axisymmetric contribution of course, we end up with:
\begin{align}
\nonumber\label{p9_potential_modes}
 & \Phi_{\text{P9}}(r,\theta,\phi) = A_0^0 Y_0^0 \cos\psi_0^0 +A_1^1Y_1^1\cos\psi_1^1 +A_2^0Y_2^0\cos\psi_2^0 \\\nonumber
  &+A_2^2Y_2^2\cos\psi_2^2
  +A_3^1Y_3^1\cos\psi_3^1 +A_3^3Y_3^3\cos\psi_3^3\\\nonumber
  &=-\frac{1}{\sqrt{4\pi}}A_0^0 -\sqrt{\frac{3}{8\pi}}\sin\theta\cos\phi A_1^1 +\sqrt{\frac{5}{16\pi}}(3\cos^2\theta -1) A_2^0 \\\nonumber
  &+\sqrt{\frac{15}{32\pi}}\sin^2\theta \cos2\phi\cos\psi_2^2 A_2^2+\sqrt{\frac{35}{64\pi}}\sin^3\theta\cos3\phi \cos\psi_3^3 A_3^3  \\
  &  -\sqrt{\frac{21}{64\pi}}\sin\theta(5\cos^2\theta-1)\cos\phi\cos\psi_3^1 A_3^1.
\end{align}
Here, mode amplitudes and phases are interpolated over $r$, the test particle's orbital radius. In order to recover the secular perturbation of P9 up to the specified order, one has to then average Eq.\ref{p9_potential_modes} numerically over the test particle's orbital angle, and a range of eccentricities, for any desired semi-major axis. In so doing, one first expresses harmonics in terms of orbital elements using
\begin{align}
\nonumber
    x=r\sin\theta\cos\phi = r [&\cos(\Omega-\varpi_{p9})\cos(\omega+f)- \\
    &\sin(\Omega-\varpi_{p9})\sin(\omega+f)\cos (i)]
\end{align}
\begin{align}\nonumber
    y=r\sin\theta\sin\phi =r &\sin(\Omega-\varpi_{p9})\cos(\omega+f)+\\
    &\cos(\Omega-\varpi_{p9})\sin(\omega+f)\cos (i)]
\end{align}
\begin{align}
    z&=r\cos\theta =r\sin(\omega+f)\sin(i).
\end{align}
Then, one isolates functions of the true anomaly $f$, which are numerically averaged to yield: 
\begin{equation}\label{flmn}
    f_{l,m,n}(a, e) = \langle A_l^m(r) \cos(n f)\rangle
\end{equation}
with $\langle\hspace{.1cm}\rangle$ standing for averaging over a test particle's orbit. Samples of the averaged functions $f_{l,m,n}(a,e)$ are shown in Fig.\ref{modes_fig}. We then transform from orbital elements to the vector notation using the definition of the angular momentum and eccentricity vectors 
\begin{equation}
\vec{j}= \sqrt{1-e^2} \begin{pmatrix}
\sin i \sin\Omega 	 \\ 
l-\sin i \cos\Omega \\
\cos i
\end{pmatrix}
\end{equation}
\begin{equation}
\vec{e}=e \begin{pmatrix}
\cos\omega\cos\Omega-\sin\omega\cos i \sin\Omega\\
\cos\omega\sin\Omega+\sin\omega\cos i \cos\Omega\\
\sin\omega\sin i
\end{pmatrix}.
\end{equation}
Using $\vec{e}=e {\hat{u}}$, and $\vec{j}=\sqrt{1-e^2} {\hat{n}}$, we can write the secular perturbation of P9 as
\begin{align}  \nonumber  \label{p9_potential_harmonics}
     \Bar{\Phi}_{\text{P9}}(a,e) = &\Upsilon_0(a,e) +\Upsilon_1(a,e)e_u+\Upsilon_2(a,e)j_n^2 +\Upsilon_3(a,e)e_n^2\\\nonumber
    &+\Upsilon_4(a,e)j_u^2+ \Upsilon_5(a,e)e_u^2 
    +\Upsilon_6(a,e)e_u j_n^2
     \\\nonumber
    &+\Upsilon_7(a,e)e_n j_u j_n
    +\Upsilon_8(a,e)e_u e_n^2+\Upsilon_9(a,e)e_u j_u ^2\\
    &+\Upsilon_{10}(a,e)e_u^3
\end{align}
with subscripts indicating the component of the particle's vectors along the basis vectors associated with the P9 plane, and with the functions $\Upsilon_i(a,e)$ given by:
\begin{align}\nonumber
    \Upsilon_0(a,e)&= -\frac{1}{\sqrt{4\pi}}f_{0,0,0}+\frac{1}{2}\sqrt{\frac{5}{16\pi}}[f_{2,0,0}-3f_{2,0,2}]\\\nonumber
    &+\frac{1}{2}\sqrt{\frac{15}{32\pi}}\Big[f_{2,2,0}-3f_{2,2,2}\Big]\nonumber
\end{align}
\begin{align}
\nonumber
    \Upsilon_1(a,e)&=\frac{1}{e}\Bigg[\sqrt{\frac{3}{8\pi}}f_{1,1,1}+\frac{1}{4}\sqrt{\frac{21}{64\pi}}\Big[-f_{3,1,1}+5f_{3,1,3}\Big]\\\nonumber&+\frac{3}{4}\sqrt{\frac{35}{64\pi}}\Big[f_{3,3,1}-5f_{3,3,3}\Big]\Bigg]
    \nonumber
\end{align}
\begin{align}\nonumber
    \Upsilon_2(a,e)&=\frac{1}{1-e^2}\Bigg[\frac{3}{2}\sqrt{\frac{5}{16\pi}}\Big[-f_{2,0,0}+f_{2,0,2}\Big]\Bigg]+\frac{1}{2}\Upsilon_4(a,e)
\end{align}
\begin{equation}
\nonumber
     \Upsilon_3(a,e)=\frac{1}{e^2}\Bigg[3\sqrt{\frac{5}{16\pi}}f_{2,0,2}\Bigg]+\frac{1}{2}\Upsilon_5(a,e)\nonumber
     \end{equation}
     \begin{equation}\nonumber
    \Upsilon_4(a,e)=\frac{1}{1-e^2}\Bigg[\sqrt{\frac{15}{32\pi}}\Big[-f_{2,2,0}+2f_{2,2,2}\Big]\Bigg]\nonumber
    \end{equation}
     \begin{equation}\nonumber
    \Upsilon_5(a,e)=\frac{1}{e^2}\Bigg[2\sqrt{\frac{15}{32\pi}}f_{2,2,2}\Big]\Bigg]\nonumber
    \end{equation}
    \begin{equation}\nonumber
    \Upsilon_6(a,e)=\frac{1}{e(1-e^2)}\Bigg[\frac{5}{4}\sqrt{\frac{21}{64\pi}}\Big[f_{3,1,1}-f_{3,1,3}\Big]\Bigg]+\frac{1}{4}\Upsilon_9(a,e)\nonumber
     \end{equation}
    \begin{equation}\nonumber
    \Upsilon_7(a,e)=2\Upsilon_6(a,e)\nonumber
      \end{equation}
     \begin{equation}\nonumber
     \Upsilon_8(a,e)=\frac{1}{e^3}\Bigg[-5\sqrt{\frac{21}{64\pi}}f_{3,1,3}+3\sqrt{\frac{35}{64\pi}}f_{3,3,3}\Bigg]\nonumber
      \end{equation}
     \begin{equation}\nonumber
     \Upsilon_9(a,e)=\frac{1}{e(1-e^2)}\Bigg[3\sqrt{\frac{35}{64\pi}}\Big[-f_{3,3,1}+f_{3,3,3}\Big]\Bigg]
      \end{equation}
     \begin{equation} \label{Phis_P9}
     \Upsilon_{10}(a,e)=\frac{1}{e^3}\Bigg[4\sqrt{\frac{35}{64\pi}}f_{3,3,3}\Big]\Bigg].
\end{equation}
\subsection{Equations of Motion}
Using Eq.\eqref{Mil_Je}, we derive the equations of motion for a test particle under the effect of the inner quadrupolar forcing from the giant planets and the outer forcing of the eccentric ring of P9, where the latter is driven by Eq.\eqref{p9_potential_harmonics}. Taking the basis vectors of P9's orbit $\{\hat{u}_o,\hat{v}_o,\hat{n}_o\}$ as the reference triad, we have:
\begin{align}
\label{p9_EOM3d_dj}
\nonumber
   L\frac{d\vec j}{dt}=&+\Gamma \frac{\vec j \cdot {\hat{n}_p}}{(1-e^2)^\frac{5}{2}}(\vec j\times{\hat{n}_p}) \\\nonumber
    &-\Big[2\Upsilon_2 j_n +2\Upsilon_6 j_n e_u +\Upsilon_7 e_n j_u\Big](\vec j\times{\hat{n}_o})\\\nonumber
    &-\Big[\Upsilon_7 e_n j_n +2\Upsilon_4 j_u +2\Upsilon_9 e_u j_u\Big](\vec j\times{\hat{u}_o}) \\\nonumber
    &-\Big[2\Upsilon_3 e_n +2\Upsilon_8 e_u e_n +\Upsilon_7 j_u j_n\Big] (\vec e\times{\hat{n}_o})\\\nonumber
    &-\Big[\Upsilon_1 + \Upsilon_6 j_n^2 +\Upsilon_8 e_n^2+2\Upsilon_5 e_u \\
    &\hspace{0.4cm}+\Upsilon_9 j_u^2+3\Upsilon_{10}e_u^2\Big] (\vec e\times{\hat{u}_o})
\end{align}
\begin{align}
\label{p9_EOM3d_de}
\nonumber
     L\frac{d\vec e}{dt}=&+\Gamma \frac{\vec j \cdot {\hat{n}_p}}{(1-e^2)^\frac{5}{2}}(\vec e\times{\hat{n}_p}) -\frac{\Gamma}{2}\frac{1-e^2-5(\vec j\cdot {\hat{n}_p})^2}{(1-e^2)^\frac{7}{2}}(\vec j\times\vec e) \\\nonumber
       &-\Big[2\Upsilon_{2}  j_{n} +2\Upsilon_{6}  j_{n} e_{u} +\Upsilon_{7}  e_{n} j_{u}\Big](\vec e\times{\hat{n}_o})\\\nonumber
    &-\Big[\Upsilon_7 e_n j_n +2\Upsilon_4 j_u +2\Upsilon_9 e_u j_u \Big](\vec e\times{\hat{u}_o}) \\\nonumber
    &-\Big[2\Upsilon_3 e_n +2\Upsilon_8 e_u e_n +\Upsilon_7 j_u j_n\Big] (\vec j\times{\hat{n}_o})\\\nonumber
    &-\Big[\Upsilon_1 + \Upsilon_6 j_n^2 +\Upsilon_8 e_n^2\\\nonumber
    &\hspace{0.4cm}+2\Upsilon_5 e_u +\Upsilon_9 j_u^2+3\Upsilon_{10}e_u^2\Big] (\vec j\times{\hat{u}_o})\\\nonumber
    &-\Big[D_0 +D_1 e_u +D_2 j_n^2 +D_3 e_n^2 +D_4 j_u^2 +D_5 e_u^2 +D_6 e_u j_n^2\\
    &\hspace{0.4cm} +D_7e_nj_uj_n+D_8 e_u e_n^2 +D_9 e_u j_u^2 +D_{10}e_u^3\Big](\vec j\times\vec e)
\end{align}
where we have dropped the explicit dependence of $\Upsilon_i$ on $a$ and $e$, and we have used $L=\sqrt{GM_\odot a}$, 
\begin{equation}
   D_i(a,e) \equiv \frac{1}{e}\frac{d\Upsilon_i(a,e)}{de},
\end{equation}
and 
\begin{equation}
    \Gamma= \frac{3}{4}\frac{G M_\odot}{a}\sum_{i=1}^{i=4} \frac{m_i a_i^2}{M_\odot a^2} 
    \label{Gamma}
\end{equation}
with the sum taken over the four gaseous giants.

When considering co-planar dynamics, the Lenz vector equation reduces to
\begin{align}\label{planar_EOM}
\nonumber
    L\frac{d\vec{e}}{dt}= \Bigg[&\Gamma \frac{e}{(1-e^2)^2} \mp \Upsilon_1l_p +2\Upsilon_2el_p - 2\Upsilon_5el_p \pm\Upsilon_6(3e^2-1)l_p\\\nonumber
    &\mp3\Upsilon_{10}e^2l_p  -D_0e l_p\mp D_1e^2l_p-D_2el_p^3- D_5e^3l_p \\
    & \mp D_6e^2 l_p^3 \mp D_{10}e^4 l_p \Bigg]\hat{v}_o
\end{align}
where $l_p= |\vec j|= \sqrt{1-e^2}$, and upper/lower signs refer to apsidally aligned/anti-aligned configurations.